\documentclass[lettersize,journal]{IEEEtran}
\usepackage{amsmath,amsfonts}
\usepackage{array}
\usepackage[caption=false,font=normalsize,labelfont=sf,textfont=sf]{subfig}
\usepackage{textcomp}
\usepackage{stfloats}
\usepackage{url}
\usepackage{verbatim}
\usepackage{graphicx}
\usepackage{cite}
\usepackage{amsmath,amssymb,amsfonts}
\usepackage{color}
\usepackage{graphicx}
\usepackage{textcomp}
\usepackage{graphicx}
\usepackage{float}
\usepackage{url}
\usepackage{longtable}
\usepackage{booktabs}
\usepackage{threeparttable}

\usepackage{multirow}
\usepackage{hyperref}
\usepackage{algorithmicx}
\usepackage[linesnumbered,ruled,vlined]{algorithm2e}
\usepackage{algpseudocode}
\usepackage{amsmath}
\begin{document}

\title{An Arbitrary-Modal Fusion Network for Volumetric Cranial Nerves Tract Segmentation}
\author{Lei Xie, Huajun Zhou, Junxiong Huang, Jiahao Huang, Qingrun Zeng, Jianzhong He, Jiawei Zhang, Baohua Fan, Mingchu Li, Guoqiang Xie, Hao Chen, Yuanjing Feng
	\thanks{This work was sponsored in part by the National Natural Science Foundation of China (No. U23A20334, U22A2040, 62002327, 62303413); Natural Science Foundation of Zhejiang Province (No. LQ23F030017); HKUST Project Fund (No. FS111). (Corresponding author: Hao Chen and Yuanjing Feng.)}
	\thanks{Lei Xie, Junxiong Huang, Qingrun Zeng, Jiahao Huang, Jianzhong He, Jiawei Zhang, and Yuanjing Feng are with the College of Information Engineering, Zhejiang University of Technology, Hangzhou, China (leix@zjut.edu.cn; superzeng@zjut.edu.cn; fyjing@zjut.edu.cn).}
	\thanks{Huajun Zhou is with the Department of Computer Science and Engineering, Hong Kong University of Science and Technology, Hong Kong, China (csehjzhou@ust.hk).}
	\thanks{Hao Chen is with the Department of Computer Science and Engineering, Department of Chemical and Biological Engineering and Center for Aging Science , Hong Kong University of Science and Technology, Hong Kong, China (jhc@cse.ust.hk).}
	\thanks{Mengjun~Li is with Department of Radiology, Xiangya Hospital, Central South University, Changsha 410000, China.}
	\thanks{Guoqiang Xie is with the Department of Neurosurgery, Nuclear Industry 215 Hospital of Shaanxi Province, Xianyang, China.}
	\thanks{Baohua Fan is with the Department of Neurosurgery, Taihe Hospital of Wannan Medical College, Taihe, China.}}

\markboth{}%
{Shell \MakeLowercase{\textit{et al.}}: A Sample Article Using IEEEtran.cls for IEEE Journals}


\maketitle

\begin{abstract}
	The segmentation of cranial nerves (CNs) tract provides a valuable quantitative tool for the analysis of the morphology and trajectory of individual CNs. Multimodal CNs tract segmentation networks, e.g., CNTSeg, which combine structural Magnetic Resonance Imaging (MRI) and diffusion MRI, have achieved promising segmentation performance. However, it is laborious or even infeasible to collect complete multimodal data in clinical practice due to limitations in equipment, user privacy, and working conditions. In this work, we propose a novel arbitrary-modal fusion network for volumetric CNs tract segmentation, called CNTSeg-v2, which trains one model to handle different combinations of available modalities. Instead of directly combining all the modalities, we select T1-weighted (T1w) images as the primary modality due to its simplicity in data acquisition and contribution most to the results, which supervises the information selection of other auxiliary modalities. Our model encompasses an Arbitrary-Modal Collaboration Module (ACM) designed to effectively extract informative features from other auxiliary modalities, guided by the supervision of T1w images. Meanwhile, we construct a Deep Distance-guided Multi-stage (DDM) decoder to correct small errors and discontinuities through signed distance maps to improve segmentation accuracy. We evaluate our CNTSeg-v2 on the Human Connectome Project (HCP) dataset and the clinical Multi-shell Diffusion MRI (MDM) dataset. Extensive experimental results show that our CNTSeg-v2 achieves state-of-the-art segmentation performance, outperforming all competing methods. 
\end{abstract}


\section{Introduction}\label{sec:introduction}
\IEEEPARstart{C}{ranial} Nerves (CNs) play a crucial role in human sensations of hearing, smell, vision, and taste, as well as in non-verbal communication of our emotions through facial expressions~\cite{yoshino2016visualization}. CNs are delicate structures that require careful handling during neurosurgical procedures to avoid potential complications, which can have a significant impact on a person’s quality of life~\cite{hodaie2010vivo,jacquesson2019overcoming}. CNs tract segmentation enables visualization of CNs spatial relations with proximate entities like tumors or lesions, crucial for preoperative diagnosis and treatment strategizing~\cite{sultana2017mri,xie2023cntseg}.

Typically, CNs tract segmentation can be performed manually by selecting streamlines of represent CNs tract~\cite{zolal2016comparison,jacquesson2019probabilistic,xie2024anatomy,hu2024preoperative}. 
The strategy is based on the Regions-Of-Interest (ROIs) selection, where trained experts select CNs in an interactive way by placing ROIs~\cite{he2021comparison}. 
Alternatively, CNs tractography atlases automate the grouping of fiber streamlines into anatomically defined tracts without expert ROIs input~\cite{zeng2023automated,huang2022automatic,2020Creation,zeng2021automated}. However, the strategy of streamline selection is demanding on the whole process of diffusion tractography algorithm, which mainly includes raw Diffusion Weighted Imaging (DWI) signal extraction, Fiber Orientation Distribution (FOD) estimation, direction tracking, and fiber bundle identification.

Recently, volumetric analysis has emerged as an effective strategy for CNs tract segmentation from different MRI modalities, i.e., T1w images, T2w images, FA images, Directionally Encoded Color (DEC) images, and FOD Peaks images, which directly classifying voxels based on associated fiber bundles, bypassing traditional streamline analysis.
For example, Sultana et al. \cite{sultana2017mri} utilized deformable 3D contour models and surfaces from T2w images for direct CNs segmentation at the voxel level. Ronneberger et al.~\cite{wasserthal2018tractseg} designed TractSeg to perform volumetric white matter tract segmentation from FOD Peaks images. Similarly, AGYnet~\cite{avital2019neural} was a multimodal fusion framework for nerve segmentation using Ynet network from T1w images and DEC images. MMFnet~\cite{xie2023deep} was designed for the visual neural pathway segmentation, leveraging T1w images and FA images in a multimodal fusion network. Recent advancements, notably by CNTSeg~\cite{xie2023cntseg}, have pioneered the integration of multimodal fusion in CNs tract segmentation, leveraging structural and diffusion MRI without relying on tractography, ROI placement, or clustering. Specifically, CNTSeg enhances segmentation accuracy by fusing T1w images, Fractional Anisotropy (FA) images, and FOD Peaks images to segment five major CNs pairs directly (i.e. optic nerve CN II, oculomotor nerve CN III, trigeminal nerve CN V, and facial–vestibulocochlear nerve CN VII/VIII). However, multimodal fusion methods for CNs tract segmentation still have some problems in clinical practice. First, not all modalities are commonly available in clinical settings, leading to incomplete multimodal data. Especially with diffusion MRI, it may be difficult to collect data due to factors such as decreased image quality and patient movement. Second, existing multimodal neural segmentation methods are typically optimized for a specific modality pair, making them difficult to adapt to other modality combinations. 

\begin{figure}[]
	\centering
	\includegraphics[width=0.5\textwidth]{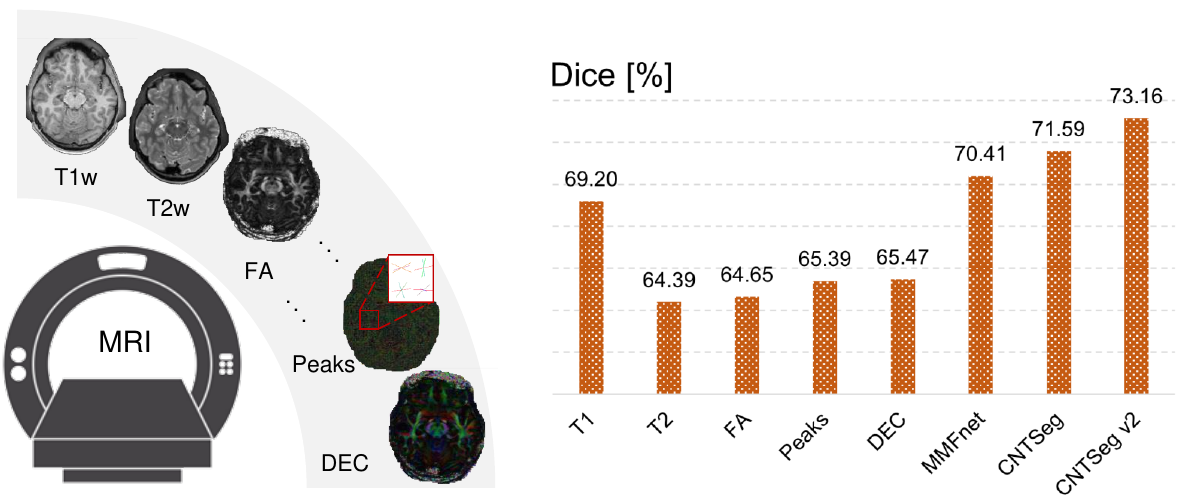}
	\caption{Comparing different single modalities (i.e., T1w images, T2w images, FA images, DEC images, and Peaks images), multimodal fusion networks including MMFnet~\cite{xie2023deep}, CNTSeg~\cite{xie2023cntseg}, and our CNTSeg-v2 on the HCP dataset.}
	\label{fig:1}
\end{figure}

By tackling the above issues, our goal is to construct a more general and effective multimodal model for volumetric CNs tract segmentation in clinical practice. First, the ease of acquisition and the importance of different modalities for neural segmentation are different. For example, diffusion MRI can improve the accuracy of segmentation but is difficult to acquire. On the other hand, structural MRI provides excellent contrast between different brain tissues (e.g., gray matter, white matter, and cerebrospinal fluid), but distinguishing CNs in the brainstem remains challenging. Instead of the missing modality models with random missingness for each modality, our CNTSeg-v2 model exploits the clinical knowledge of the simplicity of T1w image acquisition and its significant contribution to segmentation results (Fig.~\ref{fig:1}) by using the T1w image as the primary modality to guide the other modalities as complementary information to enhance segmentation capabilities of the model. Second, delineating the target region from the fine structures is a difficult challenge in medical imaging. For example, extracting the boundaries of CNs from white matter regions with relatively uniform gray matter distributions is prone to false positives and requires well-designed solutions to enhance boundary features. Therefore, we seek to design a multi-level decoding module with a specific tree structure to enhance the ability of model to recover boundary features during the decoding stage. By combining these advancements, we can construct an arbitrary-modal fusion network for volumetric CNs tract segmentation in clinical practice.

In this paper, we develop CNTSeg-v2, an arbitrary cross-modal fusion model designed for CNs tract segmentation. Based on clinical practice, the key design of our CNTSeg-v2 is the Arbitrary-Modal Collaboration Module (ACM), which uses T1w images as the primary modality and aims to effectively extract supplementary information from other available auxiliary modalities.  Furthermore, addressing CNs morphology complexity, we introduce a Signed Distance Map (SDM) and construct a Deep Distance-guided Multi-stage (DDM) decoder, which refines segmentation by correcting minor errors and mitigating discontinuities, thus enhancing accuracy. Extensive experimental results on  Human Connectome Project (HCP) dataset and the clinical Multi-shell Diffusion MRI (MDM) dataset demonstrate the effectiveness of our CNTSeg-v2 over state-of-the-art segmentation methods. The implementation code is available at~\href{https://github.com/IPIS-XieLei/CNTSeg}{https://github.com/IPIS-XieLei/CNTSeg}.

The main contributions are summarized as follows:
\begin{itemize}
	\item We propose CNTSeg-v2, an arbitrary-modal fusion network for CNs tract segmentation, which encompasses an Arbitrary-Modal Collaboration Module (ACM) designed to effectively extract supplementary information from other available auxiliary modalities.
	\item We introduce the Deep Distance-guided Multi-stage (DDM) decoder to improve the ability to recover features using the signed distance map.
	
	\item Extensive experimental results show that our CNTSeg-v2 achieves state-of-the-art segmentation performance on the HCP dataset and the clinical MDM dataset.
\end{itemize}

The rest of the paper is organized as follows. Section~\ref{sec:rw} reviews the related literature and Section~\ref{sec:METHODOLOGY} describes the overview of our CNTSeg-v2. In Section~\ref{sec:Experiments ans Results}, the experimental results on the HCP dataset and the private dataset are presented. Section~\ref{sec:Discussion} discusses the results and future works, and Section~\ref{sec:Conclusion} gives a summary of the proposed work. 
\section{Related Work}\label{sec:rw}
\subsection{CNs Tract Segmentation}\label{sec:CNs}
Early approaches~\cite{sultana2017mri} relied on specific deformable models to feature extraction for segmenting CNs tract. With the advancements in deep learning, there has been a shift towards the development of deep-based models for CNs Tract segmentation, which have produced remarkable results. For instance, \cite{dolz2017deep} proposed a deep learning classification scheme based on augmented enhanced features to segment organs at risk on the region of the CN II in patients with brain cancer. To effectively utilize multimodal information of structural MRI, \cite{mansoor2016deep} proposed the CN II segmentation
mechanism steered by deep learning features. Further, \cite{li2021two} designed a two-parallel-stages fusion network for the CN II segmentation to extract supplementary information from T1w images and FA images. On this basis, to address the problem of limited labeling samples, \cite{diakite2024lesen} proposed a label-efficient deep learning method for the CN II segmentation using T1w images and FA images. Unfortunately, current deep-based models for CNs segmentation predominantly focus on the relatively large CN II. In previous work~\cite{xie2023cntseg}, we introduced multimodal fusion network CNTSeg for CNs tract segmentation by combining structural MRI and diffusion MRI, which achieved the first segmentation of five pairs of major CNs, especially for the complex structure of the CN III and the CN V. While promising results have been achieved, the efficient fusion of multimodal data to enhance segmentation performance warrants further investigation.
\subsection{Multi-Modal Learning}\label{sec:Multi-Modal}
In clinical practice, limitations in equipment, user privacy, and working conditions pose challenges to collecting complete modality data. To mitigate this issue, arbitrary-modal fusion in multi-modal learning presents a promising approach. More recently, \cite{zhang2023cmx} introduced CMX, a unified network for RGB- and X-modal segmentation that facilitates multi-level cross-modal interactions. Similarly, \cite{wang2023multistage} developed a network for HSI-X image classification to leverage complementary information across modalities. Moreover, \cite{zhang2023delivering} introduced a universal arbitrary-modal semantic segmentation framework CMNeXt, which dynamically selects informative features from all modality-sources. 

Existing arbitrary-modal fusion methods essentially involve dedicated training strategies defined by a series of independent models, each specifically trained for different combinations of modalities, resulting in high complexity. In this paper, we propose CNTSeg-v2, an arbitrary-modal fusion network for CNs tract segmentation, which is a non-dedicated training strategy that trains one model to handle different combinations of modalities. Unlike strategies that missing modality models
~\cite{ding2021rfnet,zhang2022mmformer} by randomly deletes each modality, our CNTSegv2 is essentially a T1w- and arbitrary-modal strategy which has the ability to effectively extract supplementary information from other available auxiliary modalities by using T1w images as the primary modality, where the other auxiliary modalities can be T2w images, FA images, Directionally Encoded Color (DEC) images, and FOD Peaks images.

\section{Methodology}\label{sec:METHODOLOGY}
\subsection{Overview of CNTSeg-v2}\label{sec:Overview}
The overview of our CNTSeg-v2 is shown in Fig.~\ref{fig:CNTSeg v2}(a). We designed two parallel branches to extract features from T1w images and the other auxiliary modalities, which may include T2w images, FA images, DEC images, and FOD Peaks images, etc. Such a design is mainly based on clinical practice that T1w images are usually chosen as the main routine clinical examination for the diagnosis of CNs diseases. Moreover, the demonstration~\cite{mansoor2016deep,dolz2015fast} shows that T1w images are crucial for CNs tract segmentation relative to other modalities. Our CNTSeg-v2 adopts an encoder-decoder architecture, where the encoder adopts a two-branch design to extract features efficiently, and the decoder employs a multi-stage shape enhancement module to improve segmentation performance. The two branches of the encoder correspond to the main branch for T1w images and the auxiliary branch for other auxiliary modalities. This strategy enables accurate CNs tract segmentation even when only T1w images are available, while efficiently extracting complementary features of other available auxiliary modalities. 
Specifically, the two branches of the encoder involve the processing of input modality in a parallel but interactive manner, each of which is designed to capture the unique characteristics of the respective. At the same time, we design an Arbitrary-Modal Collaboration Module (ACM) between the two branches for guiding the selection of complementary informative features of the other auxiliary modalities under the T1w modality (Fig.~\ref{fig:CNTSeg v2}(b)). Furthermore, as show in (Fig.~\ref{fig:CNTSeg v2}(c)), we introduce a Dual-Attention Feature-Interactive Module (DFM) to fuse features belonging to the same level into a single feature map. After the encoder, the low-level fusion features will be forwarded to the decoder for the segmentation prediction. In the decoder, we incorporate a Deep Distance-guided Multi-stage (DDM) decoder into the original design to strengthen the shape information in the features by supervising the Signed Distance Map (SDM).

\subsection{Arbitrary-Modal Collaboration Module}\label{sec:Hub}
\begin{figure*}[]
	\centering
	\includegraphics[width=0.95\textwidth]{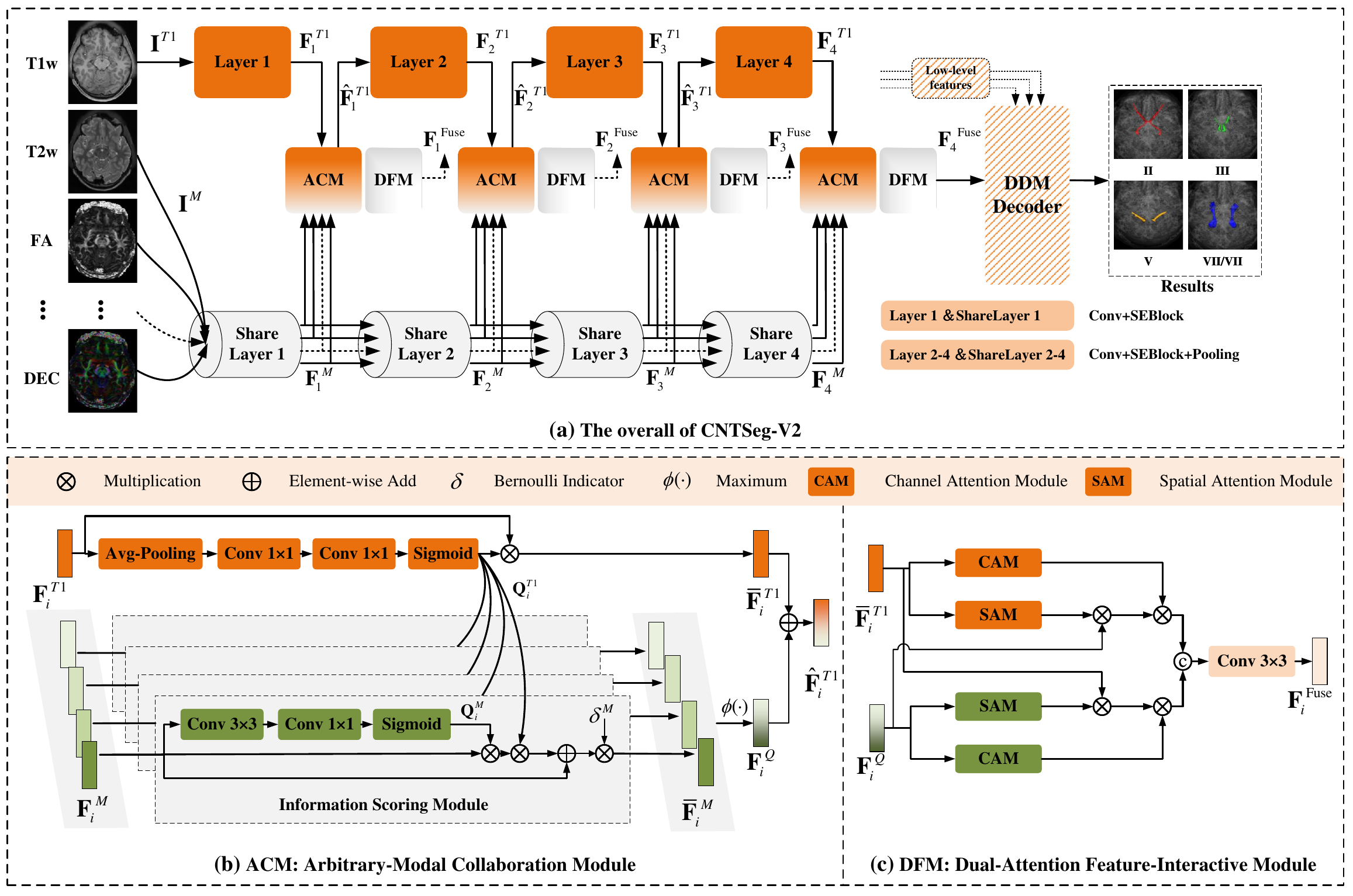}
	\caption{Overview of our CNTSeg-v2. The encoder is two-branch, which correspond to the primary branch for T1w images and the secondary branch for other other modalities (i.e., T2w, FA, Peaks, and DEC). The decoder is divided into two branches: Deep Distance-guided Multi-stage (DDM) branch and semantic supervision branch. The ACM selects informative supplementary features from other auxiliary available modalities. The input T1w feature ${\mathbf{F}_i}^{T1}$ and each complementary feature $\left\{{\mathbf{F}_i}^{M}, M \in \left\{ {\mathrm{T2},\mathrm{FA},\mathrm{Peaks},\mathrm{DEC}} \right\}\right\}$ are sent to the information scoring module to get the fused features ${\mathbf{\hat F}_i}^{T1}$. The Dual-Attention Feature-Interactive Module (DFM) leverages both spatial and channel attention to aggregate multi-modal features.}
	\label{fig:CNTSeg v2}
\end{figure*}

To perform arbitrary-modal fusion, the Arbitrary-Modal Collaboration Module (ACM) is a crucial design to extract the complementary feature of other auxiliary modalities, which is guided by the T1w modality for feature selection. As shown in Fig.~\ref{fig:CNTSeg v2}(a), the T1w images ${\mathbf{I}^{T1}} \in \mathbb{R}^{H\times W\times 1}$ is gradually processed by ${\bf{Layer}} _ {1}$ to get ${\mathbf{F}_1}^{T1} \in \mathbb{R}^{H\times W\times C}$, whereas the other modalities ${{\bf{I}}^M} \in \mathbb{R}^{H\times W\times \hat C},M \in \left\{ {\mathrm{T2},\mathrm{FA},\mathrm{Peaks},\mathrm{DEC}} \right\},\hat C \in \left\{ {1,1,9,3} \right\}$ by ${\bf{ShareLayer}}_1$  to get ${\mathbf{F}_1}^{M}\in \mathbb{R}^{H\times W\times C}$. 
As shown in Fig.~\ref{fig:CNTSeg v2}(b), at the $i^{th} \in \left\{ {1,2,3,4} \right\}$ step, the inputs of the ACM are the feature map ${\mathbf{F}_i}^{T1}$ obtained from ${\bf{Layer}} _ i$ of the T1w modality branch and the feature maps ${\mathbf{F}_i}^{M}$ obtained from ${\bf{ShareLayer}}_i$ of the complementary modality branch, respectively. The outputs of the AFM are the fused features ${\mathbf{\hat F}_i}^{T1}$, the feature map ${\mathbf{\bar F}_i}^{T1}$, and the merged feature ${\mathbf{F}_i}^{Q}$, where the fused features ${\mathbf{\hat F}_i}^{T1}$ will be fed to ${\bf{Layer}} _ {i+1}$ of the T1w modality branch, and the feature map ${\mathbf{\bar F}_i}^{T1}$ and the merged feature ${\mathbf{F}_i}^{Q}$ will be forwarded to the DFM Module. 
For ${\bf{Layer}}_i$ and ${\bf{ShareLayer}}_i$, at $i=1$, both layers incorporate an SEBlock, with the difference that ${\bf{ShareLayer}}_i$ adds a convolution layer prior to the SEBlock to accommodate multimodal inputs efficiently; at $i>1$, the architecture of layers both consist of the max-pooling operation and the SEBlock.
The specific details of ACM are given below: we design the information scoring module to extract the features of T1w and obtain the informative score mask. On the $i^{th}$ step, given T1w feature ${\mathbf{F}_i}^{T1}$, the attention module is applied to get informative score mask $\mathbf{Q}_i^{T1}$, as:
\begin{equation}\label{eq:t1w_weights}
	\mathbf{Q}_i^{T1} = \mathrm{Sigmoid}(\mathrm{DConv}_{1 \times 1}(\mathrm{AP}(\mathbf{F}_i^{T1}))),
\end{equation}
\begin{equation}\label{eq:t1w_fu}
	{\mathbf{\bar F}_i}^{T1} =\mathbf{F}_i^{T1} \otimes \mathbf{Q}_i^{T1},
\end{equation}
where the $\mathrm{DCon}{v_{1 \times 1}}(\cdot)$ means a double convolution layer with a kernel size of $1 \times 1$ and $\rm{AP}$ denote average-pooling operations. $\mathrm{Sigmoid}(\cdot)$ and $\otimes$ denote the Sigmoid activation function and the element-wise multiplication. Then, the ACM is used to calculate the informative score ${\mathbf{Q}_i}^{M}$ of each complementary feature $\left\{{\mathbf{F}_i}^{M}, M \in \left\{ {\mathrm{T2},\mathrm{FA},\mathrm{Peaks},\mathrm{DEC}} \right\}\right\}$, as:
\begin{equation}\label{eq:m_weights}
	\mathbf{Q}_i^{M} = \mathrm{Sigmoid}(\mathrm{Conv}_{1 \times 1}(\mathrm{Conv}_{3 \times 3}(\mathbf{F}_i^{M}))),
\end{equation}
where the $\mathrm{Conv}_{3 \times 3}(\cdot)$ means a double convolution layer with a kernel size of $3 \times 3$. After obtaining ${\mathbf{Q}_i}^{M}$ and ${\mathbf{Q}_i}^{T1}$ score masks, cross-modal comparisons are performed to extract informative features ${\mathbf{\bar F}_i}^{M}$ to each auxiliary modality by:
\begin{equation}
	{\mathbf{\bar F}_i}^{M} =  (\mathbf{F}_i^M + \mathbf{F}_i^M \otimes \mathbf{Q}_i^M \otimes \mathbf{Q}_i^{T1}) \otimes \delta^M,
\end{equation}
where $\delta^M \in \left\{ {0,1} \right\}$ is a Bernoulli indicator that aims to grant robustness when some modalities are missing. Then, the merged feature ${\mathbf{F}_i}^{Q}$ is extracted from the informative feature ${\mathbf{\bar F}_i}^{M}$ of each auxiliary modality by:
\begin{equation}\label{eq:max}
	{\mathbf{F}_i}^{Q} = \phi (\{ 	{\mathbf{\bar F}_i}^{M} \left| {i \in [1,2,3,4]} \right.\} ),
\end{equation}
where $\phi(\cdot)$ is an operation to select the maximum. Finally, the fused features ${\mathbf{\hat F}_i}^{T1}$ will be fed to ${\bf{Layer}} _ {i+1}$ of the T1w modality branch by:
\begin{equation}\label{eq:t1w_fu}
	{\mathbf{\hat F}_i}^{T1} ={\mathbf{\bar F}_i}^{T1} + {\mathbf{F}_i}^{Q}.
\end{equation}

To address the issue of ambiguity among features from different modalities, we introduce a Dual-Attention Feature-Interactive Module (DFM). As shown in Fig.~\ref{fig:CNTSeg v2}(c), the feature ${\mathbf{\bar F}_i}^{T1}$ and the merged feature ${\mathbf{F}_i}^{Q}$ are forwarded to DFM module, and the output of DFM module is the shared representation ${\mathbf{F}_i}^{\mathrm{Fuse}}$. The overall process can be formulated as:
\begin{equation}\label{eq:m_weights}
	\begin{cases}
		\mathbf{F}_i^{E_1} = {\mathcal{M}_s}(\mathbf{F}_i^{\mathbf{Q}} ) \otimes {\mathcal{M}_c}(\mathbf{F}_i^{\mathbf{Q}} ) \otimes {\mathbf{\bar F}_i}^{T1}\\
		\mathbf{F}_i^{E_2} = {\mathcal{M}_s}({\mathbf{\bar F}_i}^{T1}) \otimes {\mathcal{M}_c}({\mathbf{\bar F}_i}^{T1}) \otimes \mathbf{F}_i^{\mathbf{Q}}\\
		{\mathbf{F}_i}^{\mathrm{Fuse}} = \mathrm{Conv}_{3 \times 3}(\mathrm{Concate}(\mathbf{F}_i^{E_1},\mathbf{F}_i^{E_2}))
	\end{cases},
\end{equation}
where the 1D channel attention $\mathcal{M}_c$ is used to determine what information to be involved, and the 2D spatial attention map $\mathcal{M}_s$ is used to determine which part to focus~\cite{wu2023hidanet}. These operations can be formulated as follows: 
\begin{equation}\label{eq:m_weights}
	\begin{array}{*{20}{c}}
		{{\mathcal{M}_c}(f') = \mathrm{Sigmoid} ({\rm{MLP}}({\rm{AP}}(f')) + {\rm{MLP}}({\rm{MP}}(f')))}\\
		{{\mathcal{M}_s}(f') = \mathrm{Sigmoid} (\mathrm{Conv}_{7 \times 7}(\mathrm{Concate}(\mathrm{AP}(f'),\mathrm{MP}(f'))))}
	\end{array}
\end{equation}
where $\rm{MP}$ denotes the max-pooling operations, $\rm{MLP}$ is the multi-layer perceptron, and $\mathrm{Conv}_{7 \times 7}(\cdot)$ mean the convolution layer with a kernel size of $7 \times 7$.
\subsection{Deep Distance-guided Multi-stage Decoder}\label{sec:decoder}
\begin{figure}[t]
	\centering
	\includegraphics[width=0.48\textwidth]{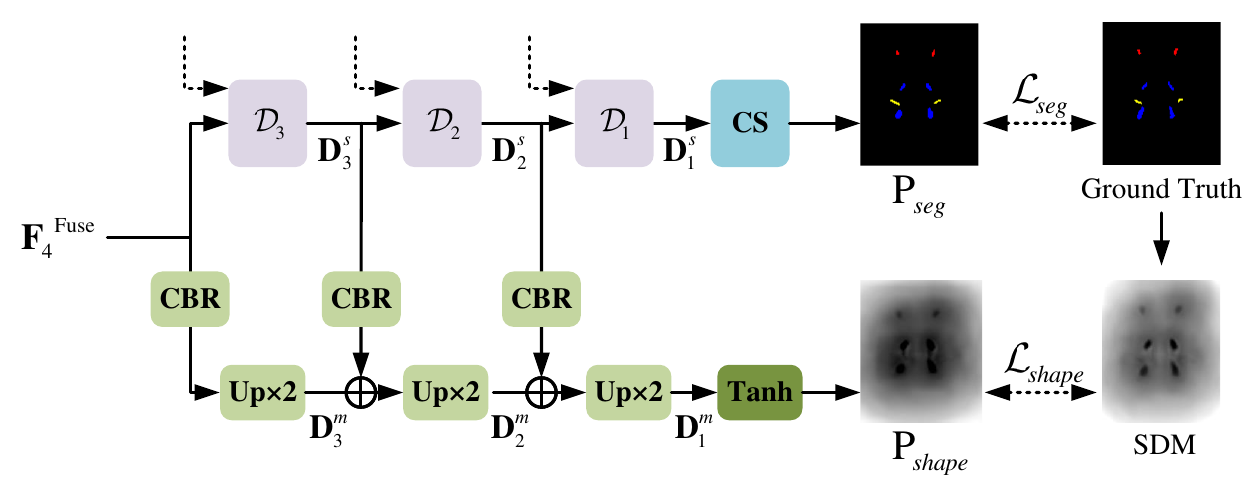}
	\caption{The overall of the Deep Distance-guided Multi-stage (DDM) decoder.}
	\label{fig:hub}
\end{figure}

As shown in Fig.~\ref{fig:hub}, to enhance the ability of the decoder to recover features, we introduce a Deep Distance-guided Multi-stage (DDM) decoder, which consists of two branches: a deep shape enhancement branch that uses the Signed Distance Map (SDM), and a original semantic supervision branch. Here we first describe how to obtain thesigned distance map, and then describe the structure of the multi-stage deep shape supervision branch. 

The signed distance of each pixel in the map denotes its proximity to the nearest object boundary, with the sign indicating its position inside (negative value) or outside (positive value) the object~\cite{wang2020deep}~\cite{yang2023exploring}. For the complex and fine structure of CNs, this framework aids neural networks in achieving accurate boundary delineation and enriching spatial relationships. Mathematically, SDM is usually formulated as: 
\begin{equation}\label{eq:sdm}
	\mathrm{S}(x) = \left\{ {\begin{array}{*{20}{c}}
			{0,x \in \partial G}\\
			{ - \mathop {\inf }\limits_{y \in \partial G} {{\left\| {x - y} \right\|}_2},x \in {G_{in}}}\\
			{ + \mathop {\inf }\limits_{y \in \partial G} {{\left\| {x - y} \right\|}_2},x \in {G_{out}}}
	\end{array}} \right.
\end{equation}
where $x$ and $y$ are different pixels in binary label. $\partial G$ denotes the contour of the target object, and $G_{in}$ and $G_{out}$ represent the internal and external regions of the target object, respectively. At the same time, a normalization process is applied, constraining its values within the range of [-1, 1].

For the original semantic supervision branch, it is composed of three decoder layer $\left\{{\mathcal{D}_{j}|j= 1,2,3}\right\}$, where each layer input is the high-dimensional feature map $h_{j}$ of the previous layer and the corresponding low-level feature map $l_{j}$ transfered from the encoder path using skip connections. The detailed operations can be formulated as:
\begin{equation}\label{eq:dj}
	{\mathcal{D}_{j}}(h_{j},l_{j}) = \mathrm{DConv}_{3 \times 3}(\mathrm{Concate}(\mathrm{TConv}(h_{j}),l_{j}),
\end{equation}
where the $\mathrm{DCon}{v_{3 \times 3}}(\cdot)$ means a double convolution layer with a kernel size of $3 \times 3$ and $\mathrm{TConv}(\cdot)$ is a transposed convolution with a kernel size of $2 \times 2$ and a stride size of 2. According to Eq.~\ref{eq:dj}, we can get the output feature map $\mathbf{D}_3^s$, $\mathbf{D}_2^s$, and $\mathbf{D}_1^s$ of each decoder layer by:
\begin{equation}\label{eq:ds}
	\begin{cases}
		{\mathbf{D}_3^s = {\mathcal{D}_3}({\mathbf{F}_4}^{\mathrm{Fuse}},{\mathbf{F}_3}^{\mathrm{Fuse}})}\\
		{\mathbf{D}_2^s = {\mathcal{D}_2}(\mathbf{D}_3^s,{\mathbf{F}_2}^{\mathrm{Fuse}})}\\
		{\mathbf{D}_1^s = {\mathcal{D}_1}(\mathbf{D}_2^s,{\mathbf{F}_1}^{\mathrm{Fuse}})}
	\end{cases}.
\end{equation}
Then, the final results $\mathbf{D}_1^s$ are classified to produce the probability map. The formula can be written as:
\begin{equation}
	{\mathrm{P}_{seg}} = \mathrm{Sigmoid}(\mathrm{Conv}_{1 \times 1}(\mathbf{D}_1^s)),
\end{equation}
where ${\mathrm{P}_{seg}}$ is the prediction result.

For the deep shape enhancement branch, it is achieved by progressively aggregating the features from each stage of the decoder and applying SDM supervision to the final output shape prediction map. Specifically, at the first stage, the feature maps ${{\bf{F}}_4}^{{\rm{Fuse}}}$ obtained by the last DFM module are transformed into $\mathrm{D}_3^m$ by the following formula:
\begin{equation}\label{eq:d3}
	{\mathbf{D}_3^m = \mathrm{U{p_2}}(\mathrm{RReLU}(\mathrm{BN}(\mathrm{Conv}_{1 \times 1}({{\bf{F}}_4}^{{\rm{Fuse}}}))))},
\end{equation}
where $\mathrm{BN}(\cdot)$, $\mathrm{RReLU}(\cdot)$, and $\mathrm{U{p_2}}(\cdot)$ represent the batchnormalization layer, the RReLU activation function, and the bilinear interpolation operation with twice the upsampling. Unlike the first stage, the second and third stages fuse the feature information $\mathbf{D}_3^s$, $\mathbf{D}_2^s$, obtained from the semantic supervision branch, while $\mathbf{D}_2^m$ and $\mathbf{D}_1^m$ can be obtained by the following formula:
\begin{equation}\label{eq:rrelu}
	\begin{cases}
		{\mathbf{D}_2^m = \mathrm{U{p_2}}(\mathrm{RReLU}(\mathrm{BN}(\mathrm{Conv}_{1 \times 1}(\mathbf{D}_3^s))) \oplus \mathbf{D}_3^m)}\\
		{\mathbf{D}_1^m = \mathrm{U{p_2}}(\mathrm{RReLU}(\mathrm{BN}(\mathrm{Conv}_{1 \times 1}(\mathbf{D}_2^s))) \oplus \mathbf{D}_2^m)}
	\end{cases}.
\end{equation}
Finally, to be in line with the value of the signed distance map, we use the Tanh activation function $\mathrm{Tanh}(\cdot)$ to get the final output from the feature map $\mathbf{D}_1^m$.  The formula is as follows: 
\begin{equation}\label{eq:tanh}
	{\mathrm{P}_{shape}} = \mathrm{Tanh}(\mathbf{D}_1^m),
\end{equation}
where $\mathrm{D}_2^m$ is the prediction result of shape supervision branch.
\subsection{Loss Function}\label{sec:Loss}
The loss function of the our CNTSeg-v2 is defined as the combination of segmentation loss $\mathcal{L}_{seg}$ and shape enhancement loss $\mathcal{L}_{shape}$, as shown in the following formula:
\begin{equation}\label{eq:ce_loss2}
	{{\mathcal{L}}_{total}} = \lambda {{\mathcal{L}}_{seg}} + (1-\lambda) {{\mathcal{L}}_{shape}},
\end{equation}
where $\lambda$ is hyper-parameters to balance the learning weights of multiple tasks, which is set to 0.5 in our experiments. 

For calculating the segmentation loss $\mathcal{L}_{seg}$, we use the Dice Loss ${\mathcal{L}}_{dl}$ and Cross Entropy ${\mathcal{L}}_{bce}$ loss, which are defined as:
\begin{footnotesize}
	\begin{equation}\label{eq:dice_loss}
		{{\mathcal{L}}_{dl}} =  1 - \frac{{2\sum\limits_{i = 1}^N {{{\mathrm{P}_{seg}}({x_i})}{{\mathrm{R}}({x_i})} + \varepsilon } }}{{\sum\limits_{i = 1}^N {{{\mathrm{P}_{seg}}({x_i})} + \sum\limits_{i = 1}^N {{{\mathrm{R}}({x_i})} + \varepsilon } } }},
	\end{equation}
\end{footnotesize}
\begin{footnotesize}
	\begin{equation}\label{eq:ce_loss}
		{{\mathcal{L}}_{bce}} = - \frac{1}{N}\left( {\sum\limits_{i = 1}^N {{{\mathrm{R}}({x_i})}\log ({{\mathrm{P}_{seg}}({x_i})}) + (1 - {{\mathrm{R}}({x_i})})\log (1 - {{\mathrm{P}_{seg}}({x_i})})} } \right)
	\end{equation}
\end{footnotesize}
where $N$ denotes the number of pixels, ${\mathrm{P}_{seg}}({x_i})$ is the prediction of network, ${\mathrm{R}}({x_i})$ is the reference data. The ${\mathcal{L}}_{ce}$ is defined as:
Therefore, the total segmentation loss is:
\begin{equation}\label{eq:ce_loss2}
	{{\mathcal{L}}_{seg}} = {{\mathcal{L}}_{dl}} + {{\mathcal{L}}_{bce}}.
\end{equation}

For calculating the shape enhancement loss $\mathcal{L}_{shape}$, we use Smooth L1 Loss to measure the effect of shape supervision of the model, as shown in the following formula: 
\begin{footnotesize}
	\begin{equation}\label{eq:ce_loss1}
		{{\mathcal{L}}_{shape}} = \left\{ {\begin{array}{*{20}{c}}
				{\frac{1}{{2N}}\sum\limits_{i = 1}^N {{{\left\| {\mathrm{S}({x_i}) - {\mathrm{P}_{shape}}({x_i})} \right\|}^2},\mathrm{if}\left| {\mathrm{S}({x_i}) - {\mathrm{P}_{shape}}({x_i})} \right| < 1} }\\
				{\frac{1}{N}\sum\limits_{i = 1}^N {\left\| {\mathrm{S}({x_i}) - {\mathrm{P}_{shape}}({x_i})} \right\|,\mathrm{otherwise}} }
		\end{array}} \right.
	\end{equation}
\end{footnotesize}
where ${\mathrm{S}_{shape}}({x_i})$ represents the signed distance map label of the corresponding pixel, and ${\mathrm{P}_{shape}}({x_i})$ represents the prediction result of the shape supervision branch.

\section{Experiments}\label{sec:Experiments ans Results}
\subsection{Experimental Settings}
\subsubsection{Dataset}\label{sec:da}
We evaluate our CNTSeg-v2 on two datasets, including the public HCP dataset~\cite{sotiropoulos2013advances,van2013wu} and the clinical MDM dataset~\cite{tong2020multicenter}, with the following detailed parameters: 

\noindent \textbf{HCP dataset}. We select 102 cases from the HCP dataset, which were scanned at Washington University in St.Louis on a customized Siemens Skyra 3T scanner (Siemens AG, Erlangen, Germany).  The parameters of dMRI data: 18 base images with b-values = 0 s/mm$^2$ and 270 gradient directions. b = 1000, 2000, and 3000 s/mm$^2$, TR = 5520 ms, TE = 89.5 ms, matrix size = 145 × 174 × 145, resolution = 1.25 × 1.25 × 1.25 mm$^3$ voxels. FA images, DEC images, and Peaks images can be calculated by MRtrix3~\cite{tournier2019mrtrix3,jeurissen2014multi} tool. Meanwhile, the acquisition parameters of T1w images and T2w images are as follows: TR = 2400 ms, TE = 2.14 ms, matrix size = 145 × 174 × 145, resolution = 1.25 × 1.25 × 1.25 mm$^3$ voxels. 

\noindent \textbf{MDM dataset}. 10 cases from the MDM dataset were chosen to evaluate our proposed method, which has the following parameters: 6 base images with b-values = 0 s/mm$^2$ and 90 gradient directions with other three b-values of 1000, 2000, and 3000 s/mm$^2$, TR = 5400 ms, TE = 71 ms, matrix size = 220 × 220 × 93, resolution = 1.5 × 1.5 × 1.5 mm$^3$ voxels. The acquisition parameters of T1w images are as follows: TR = 5000 ms, TE = 2.9 ms, matrix size = 211 × 256 × 256, resolution = 1.2 × 1 × 1 mm$^3$ voxels. 

\begin{table*}[]
	\centering
	\caption{Comparison results of our CNTSeg-v2 and SOTA cranial nerves segmentation methods on HCP dataset. }
	\label{tab:SOTA}
	\resizebox{0.97\textwidth}{!}{%
		\begin{tabular}{c||ccccc||ccccc}
			\hline\hline
			\multirow{2}{*}{Methods}  & \multicolumn{5}{c||}{Dice [\%] $\uparrow$}                                                                                             & \multicolumn{5}{c}{Jac [\%] $\uparrow$}   \\ & CN II           & CN III & CN V            & CNVII/VIII &Mean & CN II           & CN III & CN V            & CNVII/VIII  &Mean\\ \hline
			
			CNsAtlas        &82.50±4.34	&58.03±7.30	&61.98±6.48	&67.64±8.67	&67.54±4.80
			&70.43±5.79&	41.24±7.17	&45.21±6.68	&51.67±8.88&	52.14±5.01
			
			\\
			
			TractSeg       &72.18±5.58	&65.26±6.36	&61.09±7.97&	63.02±8.71	&65.39±5.23&	56.75±6.51	&48.76±6.88&	44.42±7.78	&46.56±8.72&	49.12±5.39
			
			\\\hline
			AGYnet     &84.23±2.76	&63.73±6.56	&63.63±6.38&	70.03±8.44	&70.41±4.34	&72.85±4.01	&47.08±6.78	&46.98±6.83&	54.42±8.47&	55.33±4.57
			\\
			CNTSeg-v2 (T1w+DEC)     &84.44±4.45	&64.95±6.66	&64.7±6.24	&71.59±7.56	&71.42±4.43	&73.29±5.98	&48.44±7.04	&48.12±6.62	&56.23±8.24	&56.52±4.82

			\\\hline
			MMFnet   &84.41±3.16	&66.01±5.88	&64.84±5.35	&72.51±6.04	&71.94±3.14	&73.15±4.53	&49.55±6.47	&48.20±5.71	&57.22±7.31	&57.03±3.62
			\\
			CNTSeg-v2 (T1w+FA)   &84.72±3.67	&64.99±6.71&	64.30±5.98&	72.06±6.23	&71.52±3.65	&73.65±5.14	&48.48±6.96	&47.66±6.35	&56.68±7.37&	56.62±4.12
			
			\\\hline
			CNTSeg    &84.35±2.97	&66.74±6.47	&63.69±7.49&	68.78±8.04	&71.59±4.24&	73.04±4.26	&50.42±7.03&	47.14±7.66	&52.94±8.65&	55.88±4.89
			\\
			CNTSeg-v2 (T1w+FA+Peaks)   &85.19±3.40	&65.73±6.15	&65.82±6.07	&72.28±6.95	&72.26±3.62	&74.35±4.82	&49.25±6.56	&49.35±6.65	&57.02±7.94	&57.49±4.07

			\\ \hline
			CNTSeg-v2    & \textbf{\textcolor{black}{85.74±3.30}}&	\textbf{\textcolor{black}{67.38±6.02}}	&\textbf{\textcolor{black}{66.63±5.80}}	&\textbf{\textcolor{black}{72.90±6.41}}&\textbf{\textcolor{black}{73.16±3.51}}&\textbf{\textcolor{black}{	75.18±4.70}}	&\textbf{\textcolor{black}{51.11±6.59}}&	\textbf{\textcolor{black}{50.24±6.43}}	&\textbf{\textcolor{black}{57.73±7.56}}	&\textbf{\textcolor{black}{58.56±4.04}}

			\\ \hline\hline
			
			\multirow{2}{*}{}  & \multicolumn{5}{c||}{Precision [\%] $\uparrow$}                                                                                             & \multicolumn{5}{c}{ASSD $\downarrow$}   \\ & CN II           & CN III & CN V            & CNVII/VIII &Mean & CN II           & CN III & CN V            & CNVII/VIII  &Mean\\ \hline
			CNsAtlas        &84.46±5.50&	61.59±10.47	&62.74±10.77&	72.73±12.22	&70.38±4.91	&0.208±0.118&	0.566±0.182	&0.581±0.208&	0.417±0.390&	0.443±0.162
			
			\\
			TractSeg    & 73.29±6.47&	66.38±9.06&	61.57±10.68&	66.69±11.60&	66.98±5.22	&0.365±0.197&	0.442±0.156	&0.568±0.188&	0.469±0.218&	0.461±0.123
			
			\\\hline
			AGYnet      &83.97±5.32&	63.29±8.73&	63.58±9.82&	73.22±11.98	&71.02±4.41	&0.175±0.053&	0.476±0.175&	0.544±0.226&	0.363±0.237&	0.390±0.130

			\\
			CNTSeg-v2 (T1w+DEC)   &85.22±4.52&	66.86±9.22&	63.40±10.16	&73.31±11.45&	72.20±4.58&	0.197±0.220&	0.464±0.198&	0.514±0.162&	0.353±0.247	&0.382±0.154

			\\\hline
			MMFnet     &84.02±5.61	&68.82±8.19&	63.61±9.87&	73.26±10.99&	72.43±4.49	&0.176±0.069	&0.471±0.195&0.514±0.163&	0.328±0.126&	0.372±0.084
			\\
			CNTSeg-v2 (T1w+FA)   &85.25±4.87	&67.50±9.38	&62.62±10.42	&74.47±10.59&	72.46±4.46&	0.179±0.084&	0.489±0.213&	0.522±0.158	&0.339±0.134&	0.382±0.091
			
			\\\hline
			CNTSeg & 85.56±5.22&	68.32±8.72&	65.22±9.96&	69.18±11.08&	72.07±4.17	&0.170±0.050	&0.445±0.257&	0.540±0.217	&0.370±0.152	&0.381±0.113
			\\
			CNTSeg-v2 (T1w+FA+Peaks)   &85.73±4.49	&66.43±9.37	&64.65±10.38	&74.04±11.25	&72.71±4.61	&0.167±0.077&	0.440±0.169&	0.512±0.195	&0.331±0.173&	0.363±0.103

			\\\hline
			CNTSeg-v2 & \textbf{\textcolor{black}{86.25±4.36}}	&\textbf{\textcolor{black}{69.38±9.27}}	&\textbf{\textcolor{black}{65.76±9.89}}	&\textbf{\textcolor{black}{74.50±11.55}}	&\textbf{\textcolor{black}{73.97±4.57}}	&\textbf{\textcolor{black}{0.161±0.095}}	&\textbf{\textcolor{black}{0.418±0.170}}	&\textbf{\textcolor{black}{0.492±0.190}}	&\textbf{\textcolor{black}{0.319±0.135}}	&\textbf{\textcolor{black}{0.348±0.098}}

			\\ \hline\hline
		\end{tabular}
	}
\end{table*}
\begin{figure*}[]
	\small
	\centerline{
		\begin{tabular}{@{}c@{}c@{}c@{}c@{}c@{}c@{}c@{}c@{}}
			
			\includegraphics[width=0.245\textwidth]{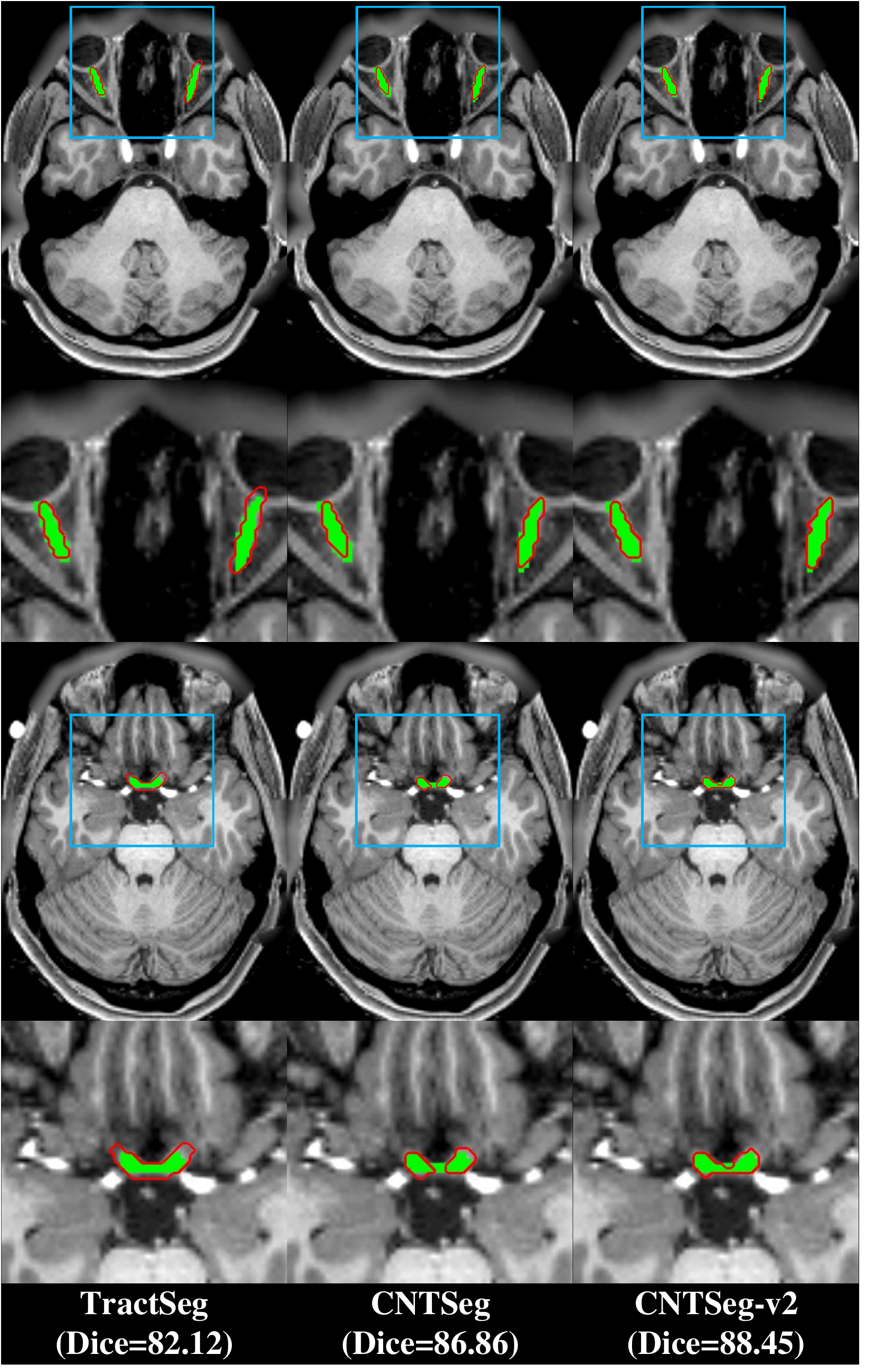}
			& \includegraphics[width=0.245\textwidth]{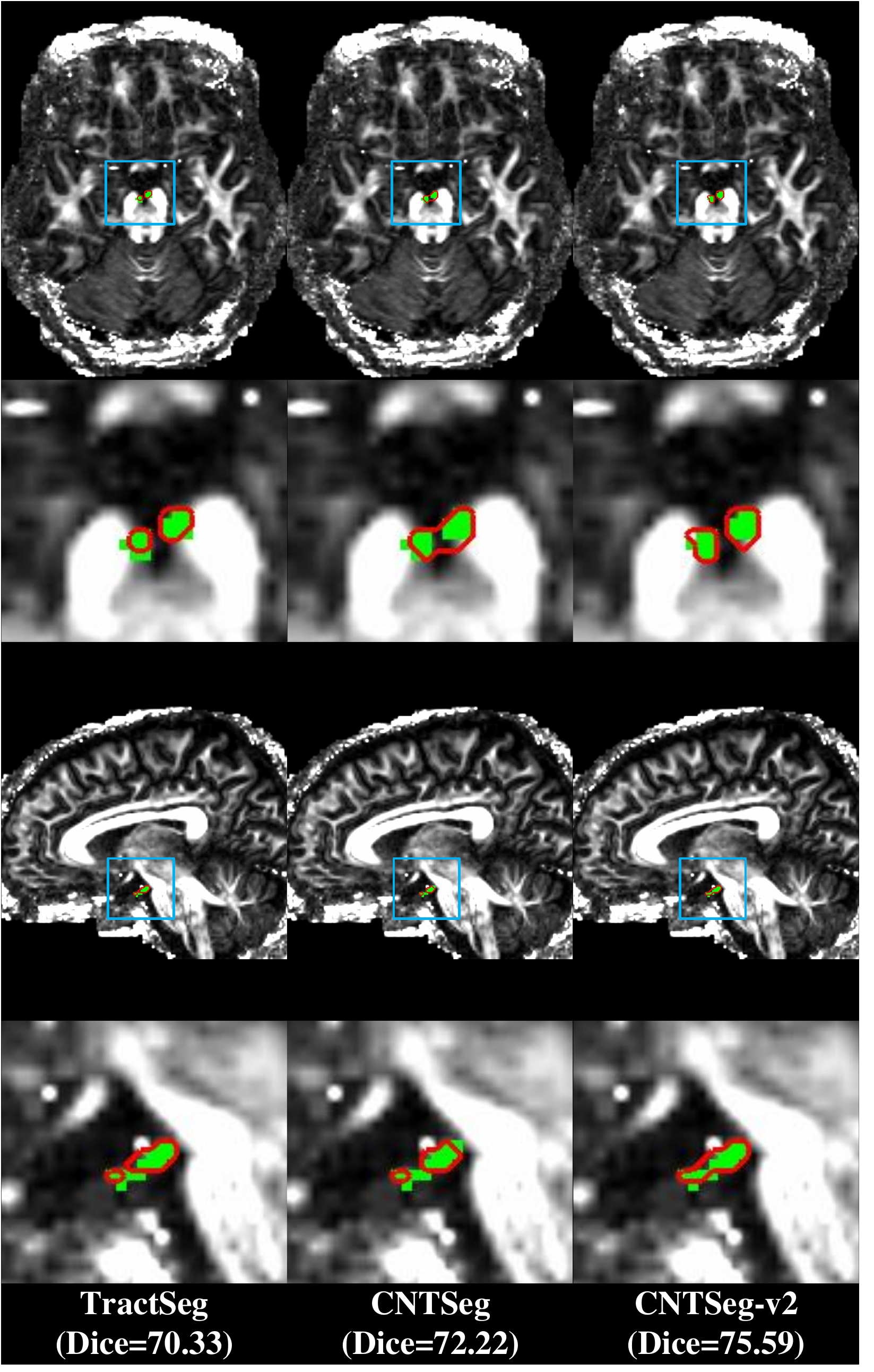}&
			\includegraphics[width=0.245\textwidth]{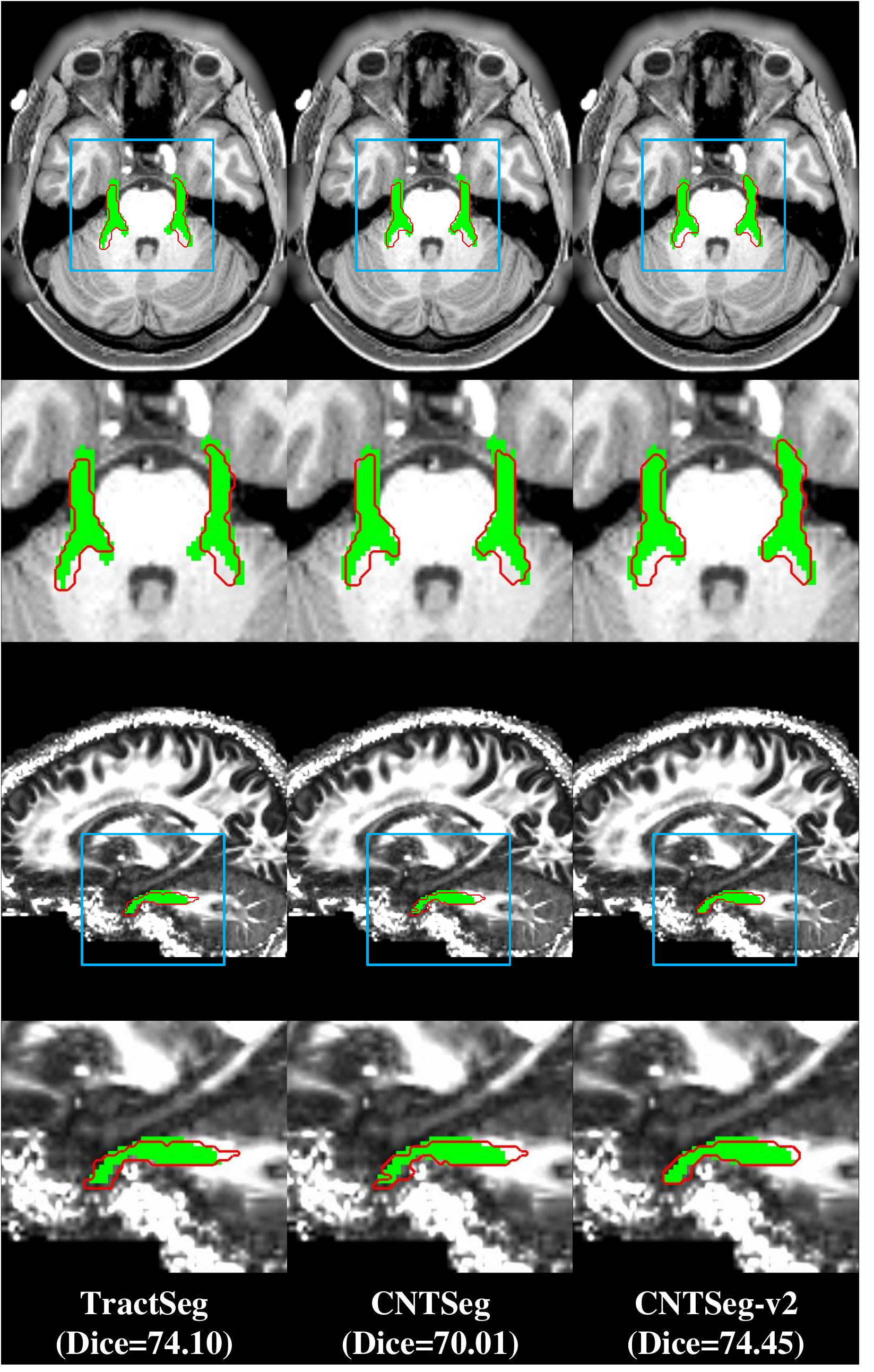}
			&\includegraphics[width=0.245\textwidth]{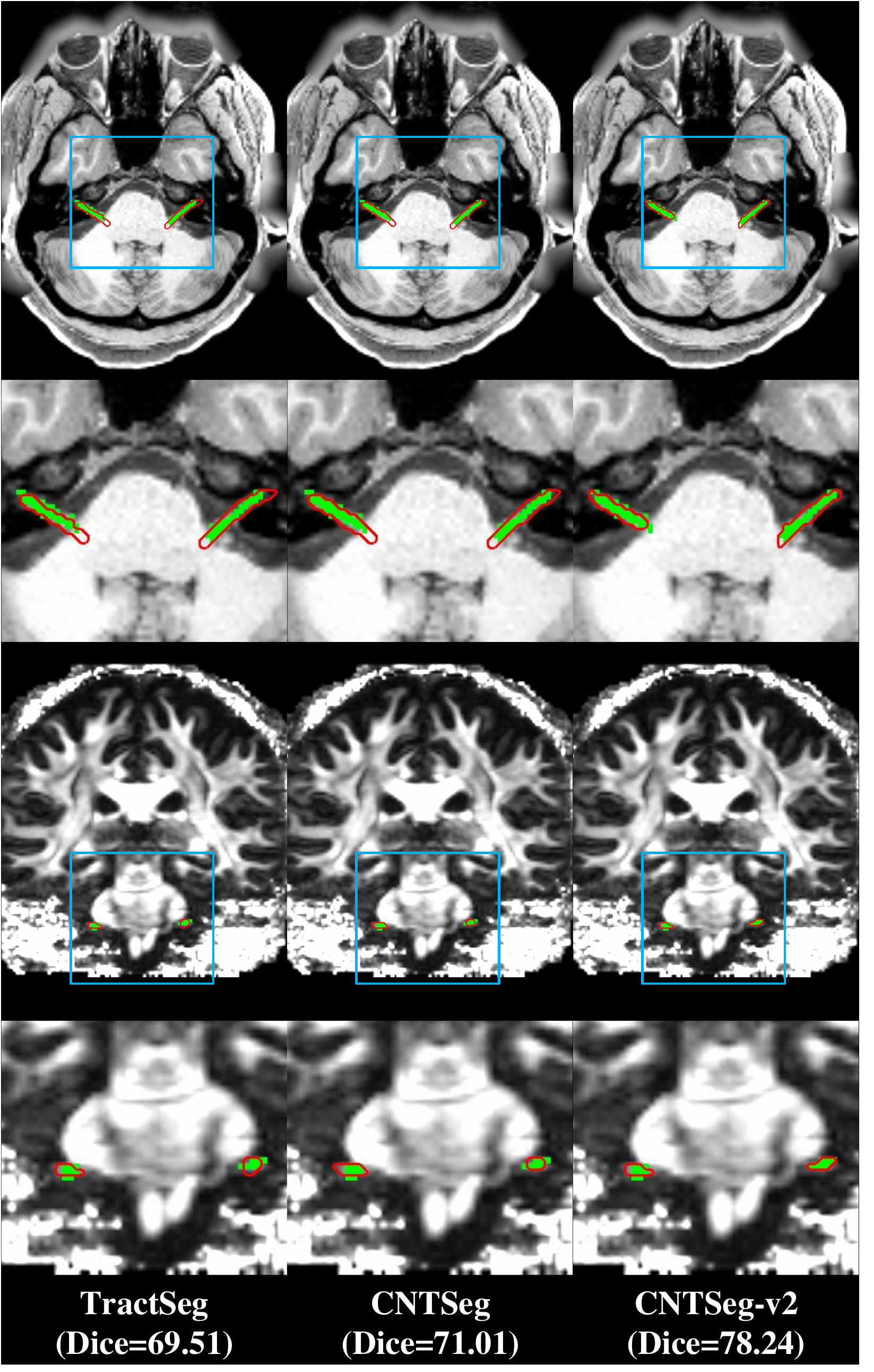}\\
			(a) CN II & (b) CN III &(c) CN V & (d) CN VII/VIII

	\end{tabular}}
	\caption{Qualitative results comparison of SOTA methods on HCP dataset. In each subfigure, the 1st and 3rd show two different slices of the segmentation results for each method in one subject, while the 2nd and 4th are the corresponding enlarged images. Green shows the reference data and red shows the segmentation of the respective method. }\label{fig:SOTA}
\end{figure*}
\subsubsection{Reference Data Generation}
For the reference data, we employ semi-automatic expert filtering to obtain binary reference data for five pairs of CNs tracts, which serve as ground truth for training and testing the proposed network~\cite{xie2023cntseg}. First, we apply multi-fiber UKF tractography to describe the 3D trajectory of the CNs. Next, we generate the CNs atlas using a fiber clustering strategy based on the HCP dataset. We then utilize this atlas for individualized CNs identification in 102 cases from the HCP dataset and 10 cases from the MDM dataset. Following this, automatic and manual screening is performed using expert-annotated precise ROIs. Finally, we map the obtained streamlines onto voxels to create binary images.
\subsubsection{Implementation Details}\label{sec:Details}
All the experiments are trained on 2 parallel NVIDIA RTX 3060 GPUs, the network are implemented using Python 3.9.7 + PyTorch 2.0.1 + Torchvision 0.15.2 + CUDA 12.1. A batch size of 32, the learning rate of 0.002 are used. We minimize the loss with an SGD optimizer using 200 epochs, and then traverse all to find the best weights. In our experiments, we employ 2D slices in the axial-plane of the volumes. The input images are processed by flipping and changing the color property randomly (i.e., brightness, contrast, and hue). Additionally, the structural  data of the HCP dataset and the MDM dataset were resized to 128 × 160 × 128 and 128 × 128 × 96 without removing any brain tissue, and the matrix size of the dMRI data are consistent with it. For 102 subjects from the HCP dataset, we perform 5-fold cross-validation with a 4:1 ratio of training to validation sets. For the MDM dataset, we also apply 5-fold cross-validation.

\subsubsection{Evaluation Metrics}
We employ four metrics to evaluate the effectiveness of all segmentation methods, i.e., Dice, Jaccard similarity index (Jac), Precision, and Average Hausdorff Distance (AHD)~\cite{wang2021annotation,xie2022semi,xie2023cntseg}. 
\begin{figure}[]
	\centering
	\includegraphics[width=0.245 \textwidth]{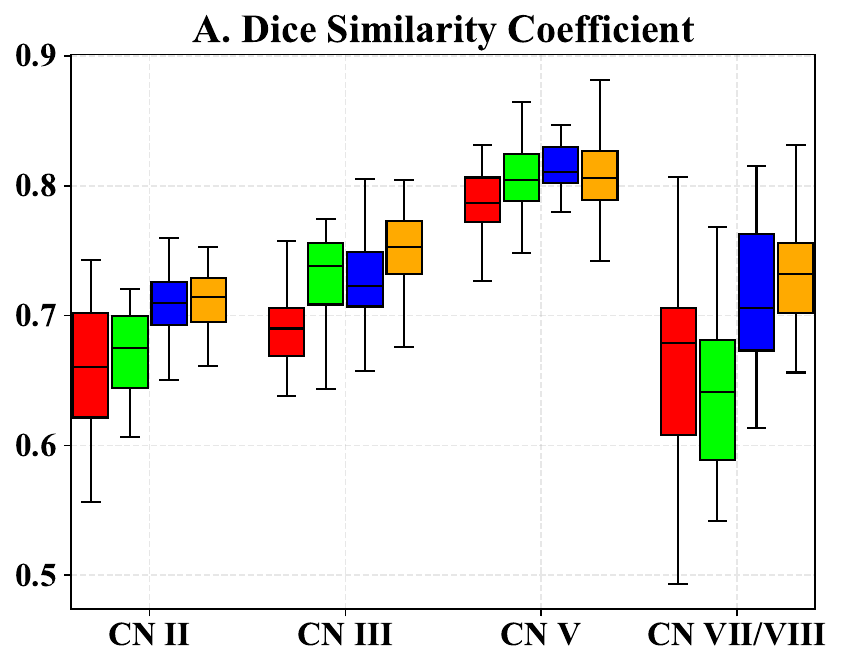}%
	\includegraphics[width=0.245 \textwidth]{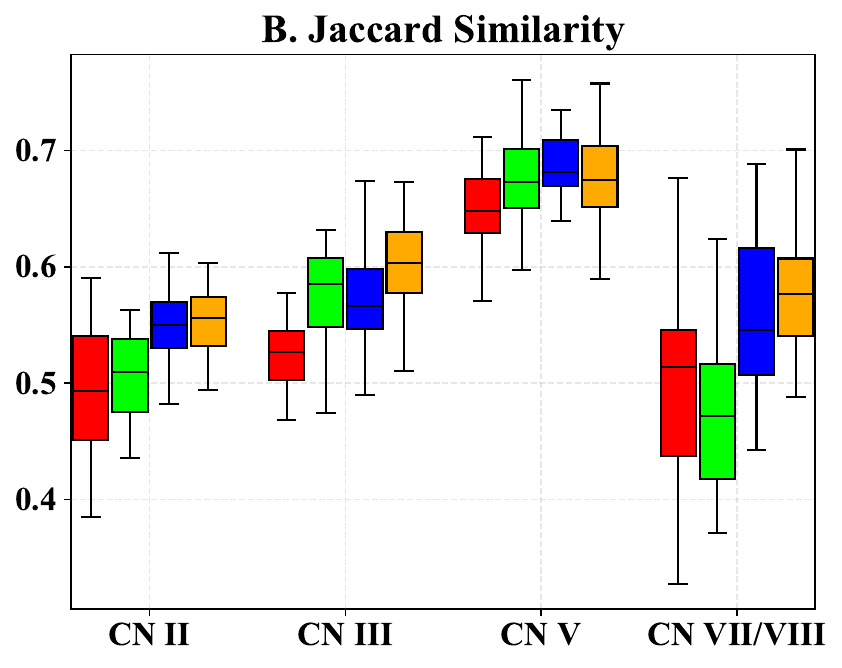}\\
	\includegraphics[width=0.245 \textwidth]{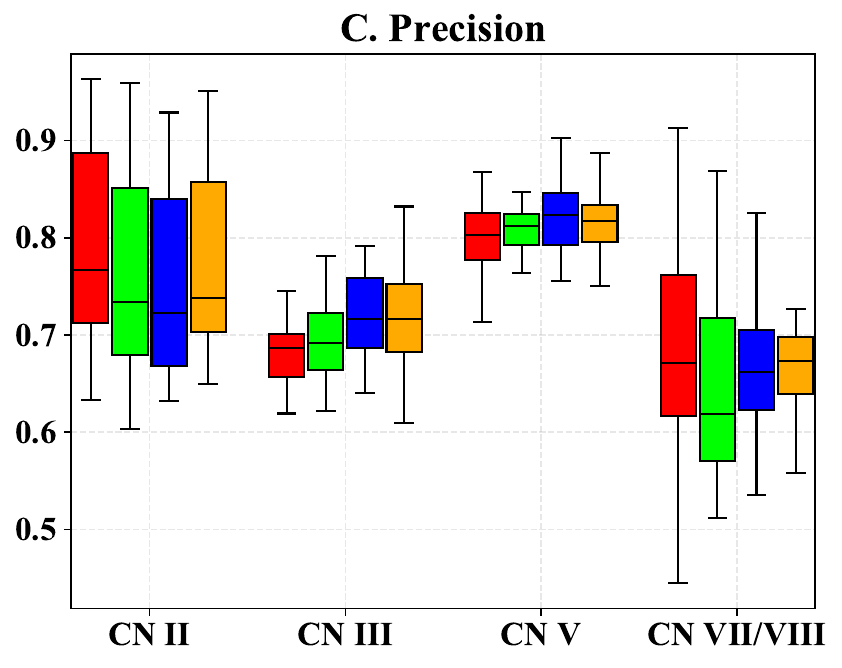}%
	\includegraphics[width=0.245 \textwidth]{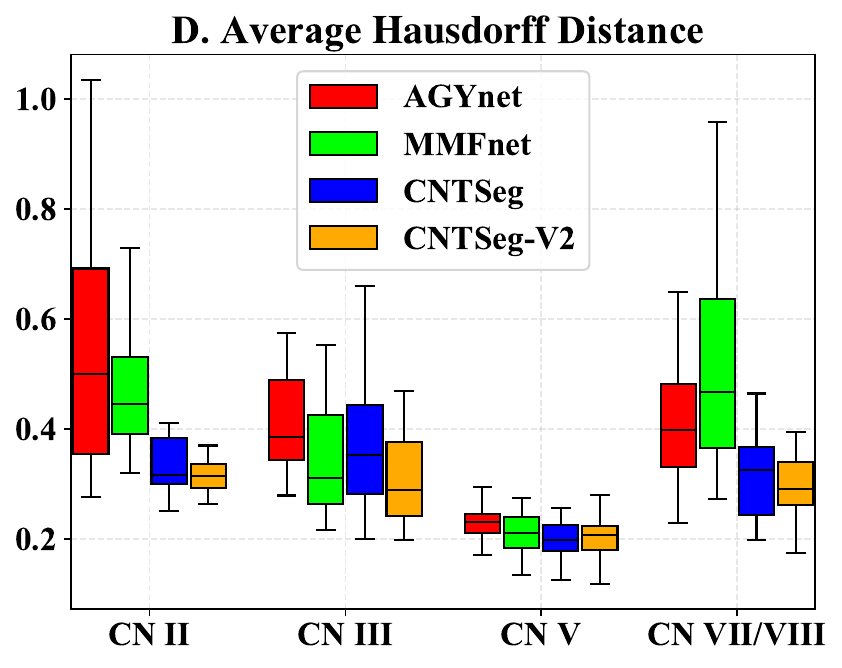}%
	\caption{Box-plots of segmentation results comparing our CNTSeg-v2 with SOTA cranial nerves segmentation methods on MDM dataset.}
	\label{fig:MDM}
\end{figure}
\begin{figure}[]
	\small
	\centerline{
		\begin{tabular}{@{}c@{}c@{}c@{}c@{}c@{}c@{}}
			
			\includegraphics[width=0.48\textwidth]{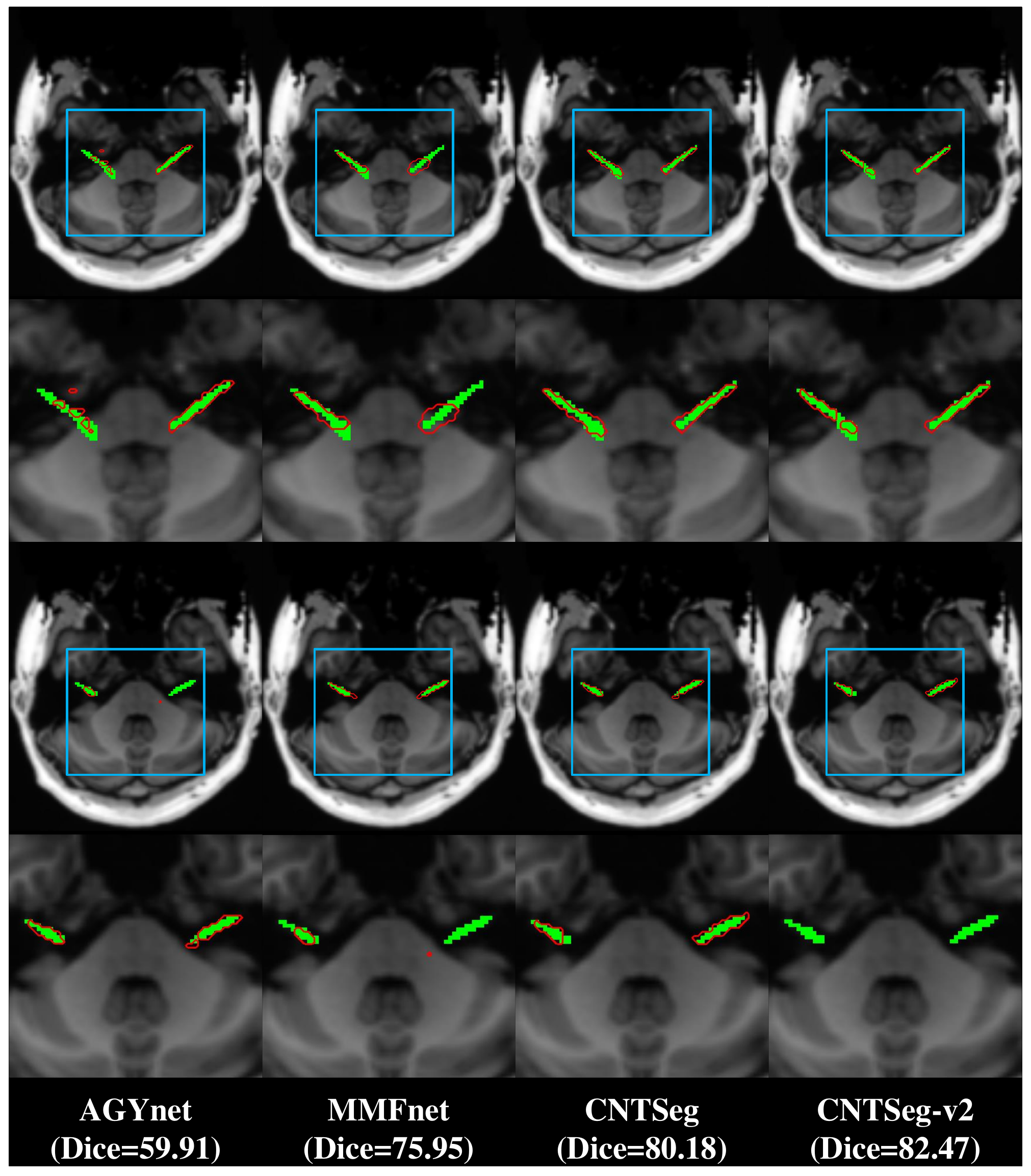}		
	\end{tabular}}
	\caption{Segmentation results of our CNTSeg-v2 on MDM dataset: Reconstruction of CN VII/VIII on MDM subject \#07. Green shows the reference tract and red shows the segmentation of the respective method. }\label{fig:mdm2}
\end{figure}
\subsection{Comparisons with SOTA Cranial Nerves Segmentation Models}
\subsubsection{Quantitative Evaluation}
To illustrate the effectiveness of our CNTSeg-v2, we conducted a comprehensive comparison with five SOTA neural segmentation methods: CNsAtlas~\cite{zeng2023automated}, TractSeg~\cite{wasserthal2018tractseg}, AGYnet~\cite{avital2019neural}, MMFnet~\cite{xie2023deep}, and CNTSeg~\cite{xie2023cntseg}. With the exception of CNTSeg, the other comparison methods are tasked with segmenting nerves rather than CNs. To ensure a fair and meaningful comparison, we implemented the segmentation of CNs in accordance with the specific modalities each method utilizes and the network architectures they employ. Specifically, TractSeg focuses on white matter tract segmentation, utilizing Peaks images fed into a CNN network. AGYnet specializes in nerve segmentation, processing T1w images and DEC images through a Ynet network. Meanwhile, MMFnet is designed for the visual neural pathway segmentation, leveraging T1w images and FA images in a multimodal fusion network. All models, including the baseline methods and our CNTSeg-v2, underwent training on identical hardware configurations and dataset splits.

Table.~\ref{tab:SOTA} give the segmentation results of our CNTSeg-v2 and the competing methods for each CNs tract in terms of Dice, Jac, Precision, and ASSD metrics. We color-code the top-1 approaches for every score in red and blue. As demonstrated in Table.~\ref{tab:SOTA}, our CNTSeg-v2 outperforms all the CNs tract segmentation models and achieves SOTA performance. For instance, our CNTSeg-v2 segments five pairs of CNs with the mean Dice, mean JAC, mean Precision, and mean ASSD of 73.16\%, 58.56\%, 73.97\%, and 0.348, which are higher than the second highest performance of MMFnet by 1.22\%, 1.53\%, 1.54\%, and 0.024, respectively. Similarly, using the same modalities as the AGYnet (\textit{T1w+DEC}), MMFnet (\textit{T1w+FA}), and CNTSeg (\textit{T1w+FA+Peaks}) for prediction, our CNTSeg-v2 model still outperforms them on all four metrics. This indicates that the effect of auxiliary modality in our designed strategy on segmentation performance improvement is effective. It is worth noting that the performance of our CNTSeg-v2 outperforms TractSeg using a single Peaks modality on four metrics 7.77\%, 9.44\%, 6.99\%, and 0.1113, respectively. The main reason is that CNs are different from WM tracts, and fusion of multimodalities can effectively improve the segmentation performance.

Moreover, we test the runtime of our CNTSeg-v2, and the overall time is basically the same as that of CNTSeg. A complete process took less than 15 minutes, which included generating the DWI, pre-processing, and generating the segmentation results. The longest runtime is for generating the DWI sequence, which takes less than 50 seconds to predict an individual when considering only network prediction.
\begin{figure}[]
	\centering
	\includegraphics[width=0.242 \textwidth]{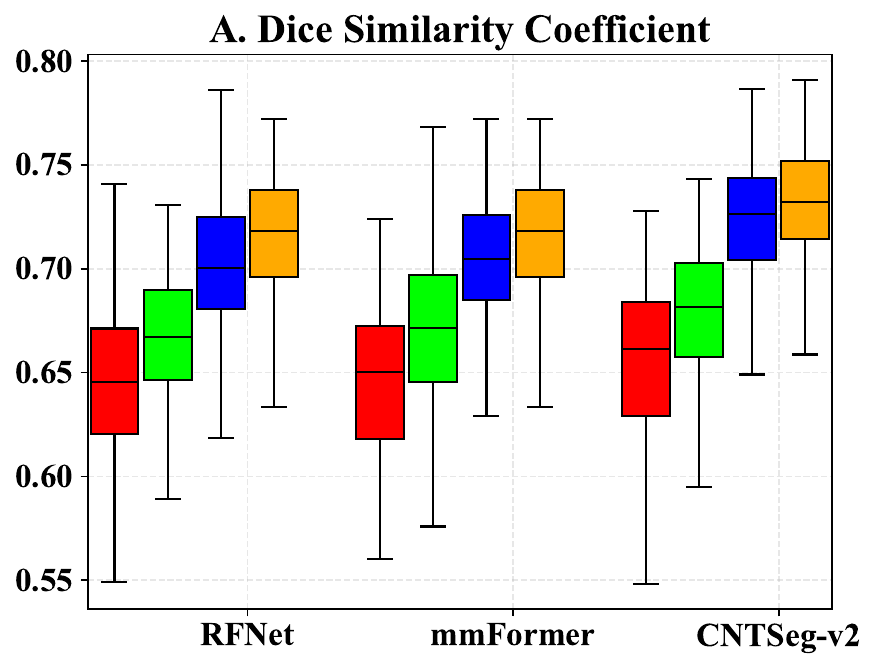}%
	\includegraphics[width=0.242 \textwidth]{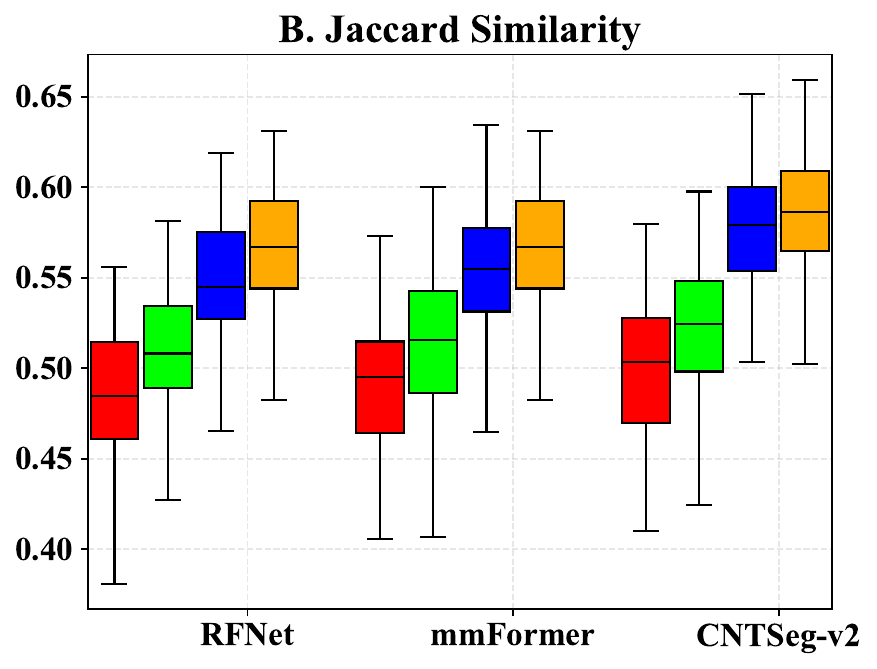}\\
	\includegraphics[width=0.242 \textwidth]{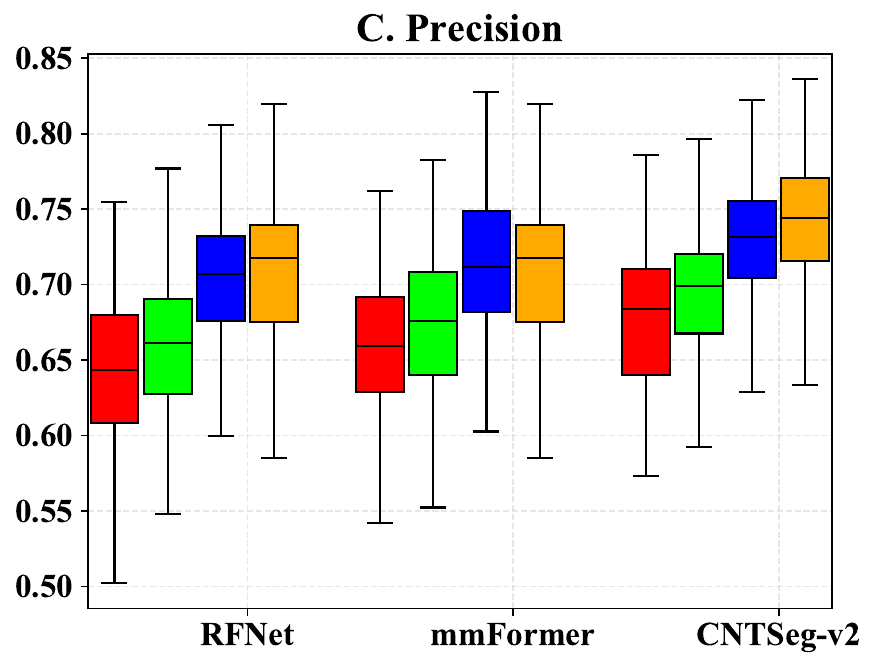}%
	\includegraphics[width=0.242\textwidth]{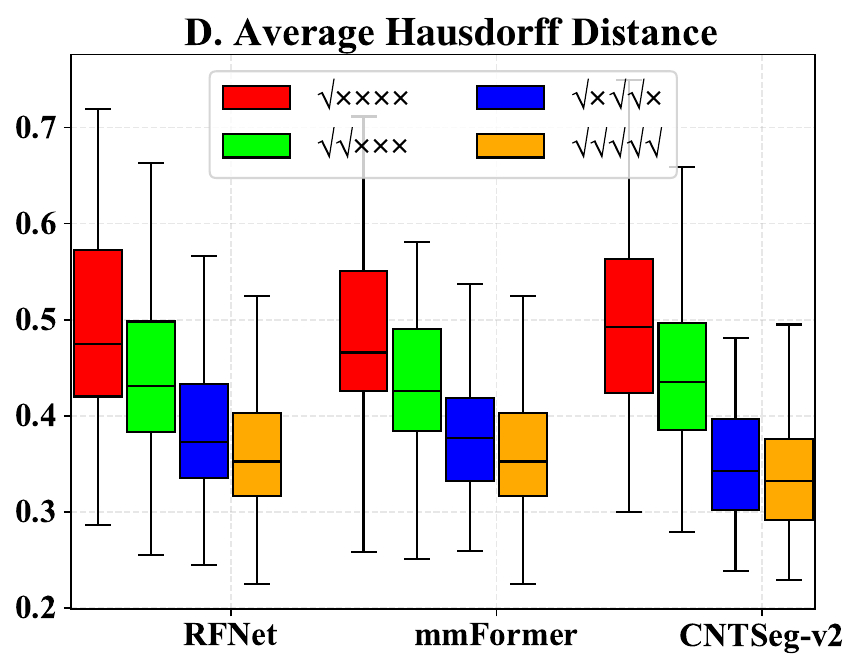}%
	\caption{Box-plots of segmentation results comparing CNTSeg-v2 with missing modality models on HCP dataset. $\checkmark$ and $\times$ represent whether T2w, FA, Peaks, and DEC images are missing, respectively.}
	\label{fig:miss1}
\end{figure}
\begin{figure}[h]
	\small
	\centerline{
		\begin{tabular}{@{}c@{}c@{}c@{}c@{}c@{}c@{}}
			
			\includegraphics[width=0.16\textwidth]{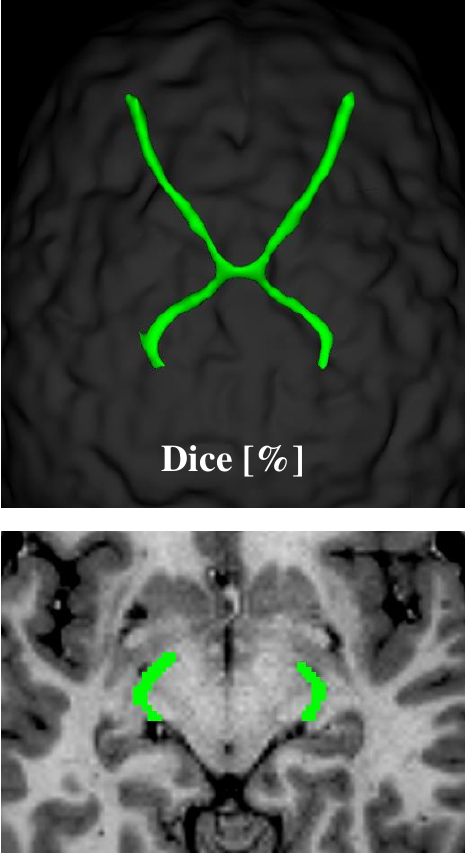}
			& \includegraphics[width=0.16\textwidth]{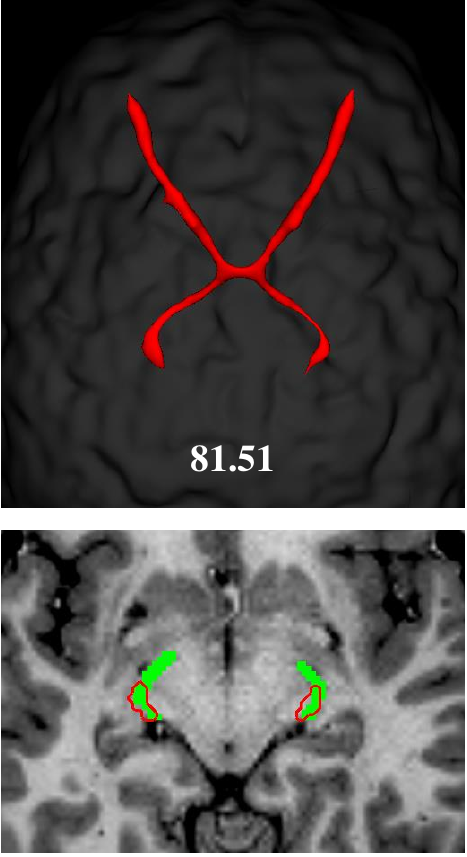}
			& \includegraphics[width=0.16\textwidth]{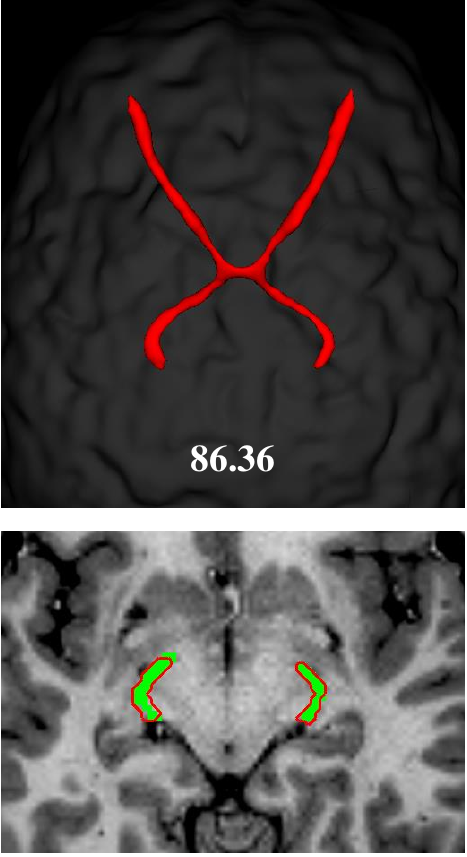}\\
			\includegraphics[width=0.16\textwidth]{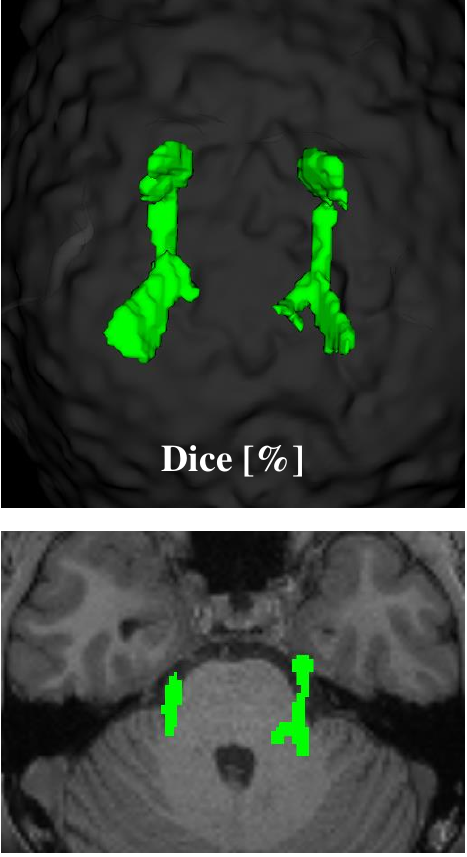}
			& \includegraphics[width=0.16\textwidth]{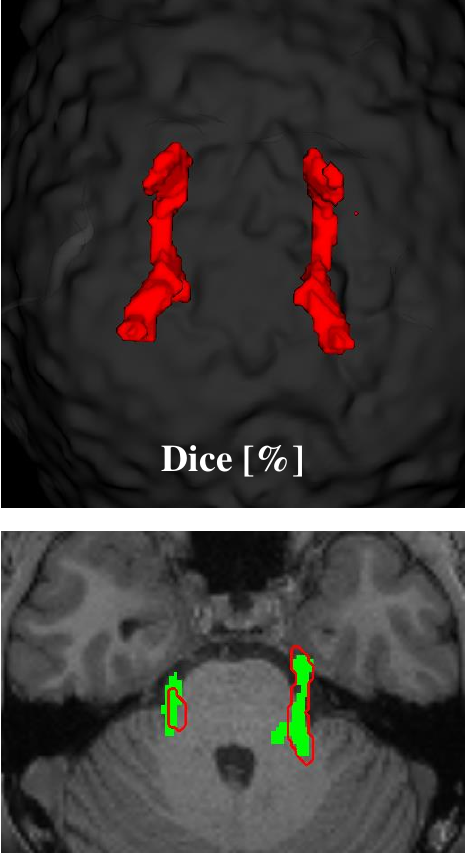}
			& \includegraphics[width=0.16\textwidth]{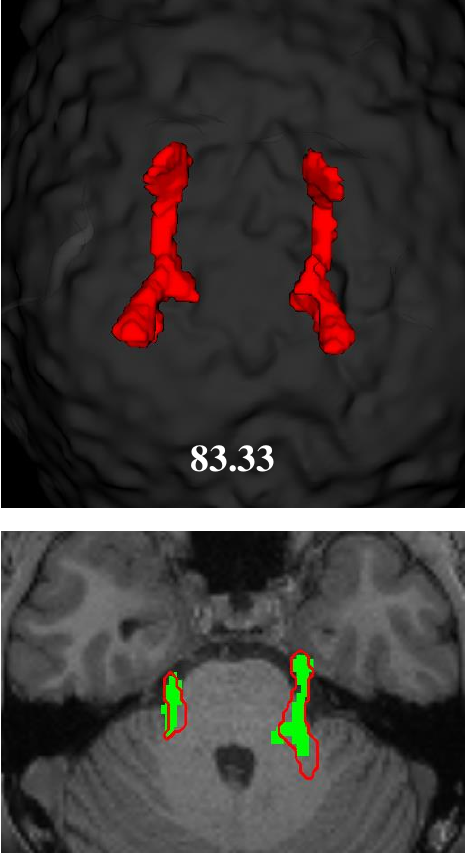}
			\\	
			(a) Reference & (b) Only T1w  &(c) All modalities 			
	\end{tabular}}
	\caption{Segmentation results of our CNTSeg-v2 with various available modalities: Reconstruction of CN II on HCP subject \#101410; Reconstruction of CN V on HCP subject \#100206. Green shows the reference tract and red shows the segmentation of the respective method.}\label{fig:miss2}
\end{figure}
\subsubsection{Qualitative Evaluation}
To demonstrate the segmentation performance of our CNTSeg-v2, we selected different CNs tract from different cases of the HCP dataset for qualitative comparison. Fig.~\ref{fig:SOTA} shows the CN II of subject \#101915, CN III of subject \#111413, CN V of subject \#119833, and CN VII/VIII of subject \#101006 in HCP dataset, respectively. As shown in Fig.~\ref{fig:SOTA}, the 1st, 3rd, 5th, and 7th rows show the 3D segmentation results of CN II, CN III, CN V, and CN VII/VIII, respectively, and the 2nd, 4th, 6th, and 8th rows are the corresponding slice images on T1w images. Green shows the reference tract and red shows the segmentation of the respective method. We can see that our CNTSeg-v2 is more complete in 3D presentation of the results relative to other competing methods, and there are significantly fewer false-positive fibers on the slice display.

\subsection{Evaluation for Different Acquisition Sources}
For a more comprehensive evaluation on model generalizability, we compare our CNTSeg-v2 with the SOTA neural segmentation methods include AGYnet, MMFnet, and CNTSeg on the MDM dataset. Fig.~\ref{fig:MDM} gives the quantitative metrics comparing our CNTSeg-v2 with the compelete methods, and we can see that our CNTSeg-v2 achieves higher performance in all four metrics. For instance, we can see that our CNTSeg-v2 segments five pairs of CNs with the mean Dice, mean JAC, mean Precision, and mean ASSD of about 75\%, 60\%, 77\%, and 0.270, which are all higher than the AGYnet, MMFnet, and CNTSeg. Fig.~\ref{fig:mdm2} shows the CN VII/VIII of subject \#07, on MDM dataset, respectively. As shown in Fig.~\ref{fig:mdm2}, the 1st and 3rd rows show two different slices of one subject, and the 2nd and 4th rows are the corresponding enlarged images. It can be seen that our CNTSeg-v2 has better coverage with anatomical regions, especially in the cisternal portion segment. These experimental results illustrate that our CNTSeg-v2 is also effective in extracting complementary information for other auxiliary modalities on the MDM dataset.
\subsection{Comparisons with Missing Modality Models}
From the point of view of missing modality, our CNTSeg-v2 allows any modalies missing except T1w images, so we compare our CNTSeg-v2 with classical missing modality models, including RFNet~\cite{ding2021rfnet}, mmFormer~\cite{zhang2022mmformer}. We apply these methods to the HCP dataset, and in order to maintain a fair comparison, we change the network to the same 2D model as ours, using five modalities for training, where T1w images were used as the modality present and the others were missing at random. Other than that, all other parameters remain at their original default settings. Fig.~\ref{fig:miss1} gives the segmentation results comparing our CNTSeg-v2 with missing modality models. $\checkmark$ and $\times$ represent whether T1w, T2w, FA, Peaks, and DEC images are missing, respectively. We can see that our CNTSeg-v2 using \textit{T1w+T2w+FA+Peaks+DEC} ($\checkmark$$\checkmark$$\checkmark$$\checkmark$$\checkmark$) achieves higher performance in all four metrics compared to only using \textit{T1w} ($\checkmark$$\times$$\times$$\times$$\times$), using structural MRI \textit{T1w+T2w} ($\checkmark$$\checkmark$$\times$$\times$$\times$) combination, using structural and diffusion MRI \textit{T1w+FA+Peaks} ($\checkmark$$\times$$\checkmark$$\checkmark$$\times$) combination. Fig.~\ref{fig:miss2} gives the qualitative comparison results of our CNTSeg-v2 using various available modalities. Quantitative and qualitative results demonstrate the superiority of our CNTSeg-v2 in dealing with the missing modality problem. For missing modality models, the missing information is supplemented by sharing the spatial mapping of the available modalities. This strategy assumes that the importance of each modality is fixed, which results in significantly inferior performance, especially when more than one modality is missing. In contrast, our CNTSeg-v2 is a T1w- and arbitrary-modal strategy that effectively utilizes complementary information from any available modalities to enhance segmentation performance while extracting features from the strongly modal T1w image.
\begin{table}[t]
	\centering
	\caption{RESULTS OF ABLATION STUDIES of our CNTSeg-v2 on HCP dataset. Bold denotes the best results. }
	\label{tab:ablation}
	\resizebox{0.5\textwidth}{!}{%
		\begin{tabular}{c||cccc}
			\hline\hline
			\multirow{2}{*}{Methods}  & \multicolumn{4}{c}{CNs}                                                                                                \\ & Dice [\%] $\uparrow$          & Jac [\%] $\uparrow$ & Precision [\%]  $\uparrow$           & ASSD $\downarrow$  \\ \hline
			
			CNTSeg-v2          & \textbf{73.16±3.51}
			&\textbf{58.56±4.04}
			& \textbf{73.97±4.57}
			& 0.348±0.098
			\\
			w/o DDM        & 73.07±2.95
			& 58.39±3.42
			& 73.91±4.25
			&\textbf{0.341±0.060}
			\\
			w/o ACM           & 72.80±3.16
			& 58.05±3.64
			&  73.68±4.77
			&  0.348±0.078
			\\
			w/o DDM+DFM+ACM           & 72.66±3.39
			& 57.93±3.89
			& 73.25±4.43
			&      0.351±0.084
			\\ \hline\hline
			
		\end{tabular}
	}
\end{table}

\begin{figure}[]
	\small
	\centerline{
		\begin{tabular}{@{}c@{}c@{}c@{}c@{}c@{}c@{}}
			
			\includegraphics[width=0.245\textwidth]{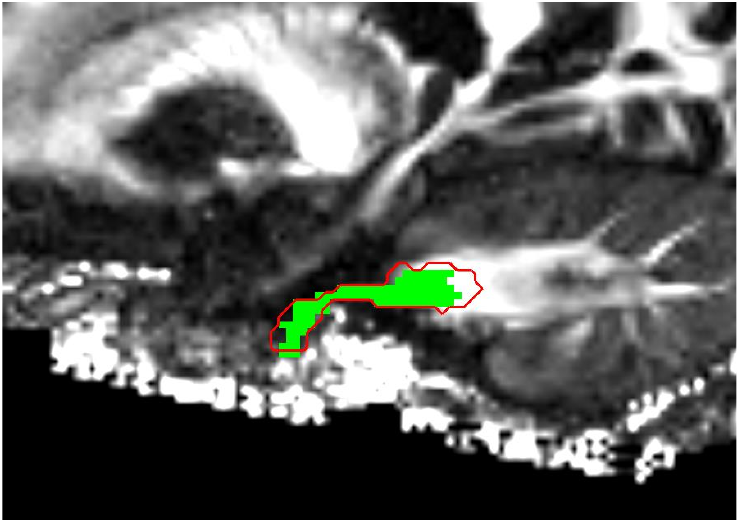}
			&\includegraphics[width=0.245\textwidth]{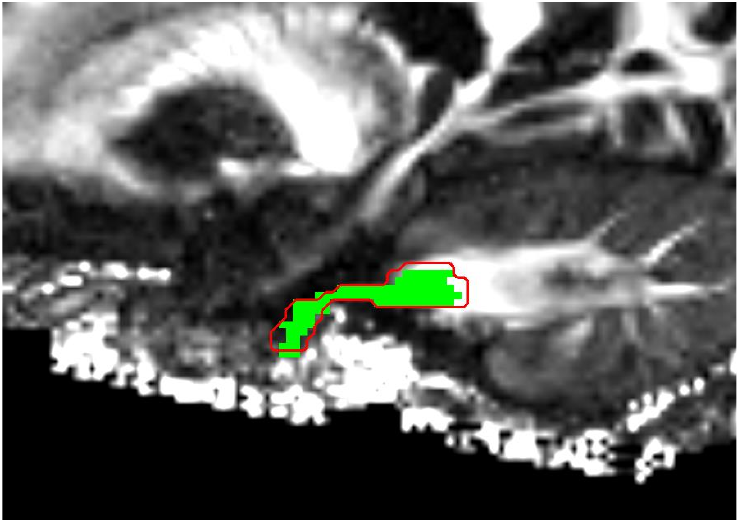}\\
			(A) w/o DDM &(B) w/o ACM\\
			\includegraphics[width=0.245\textwidth]{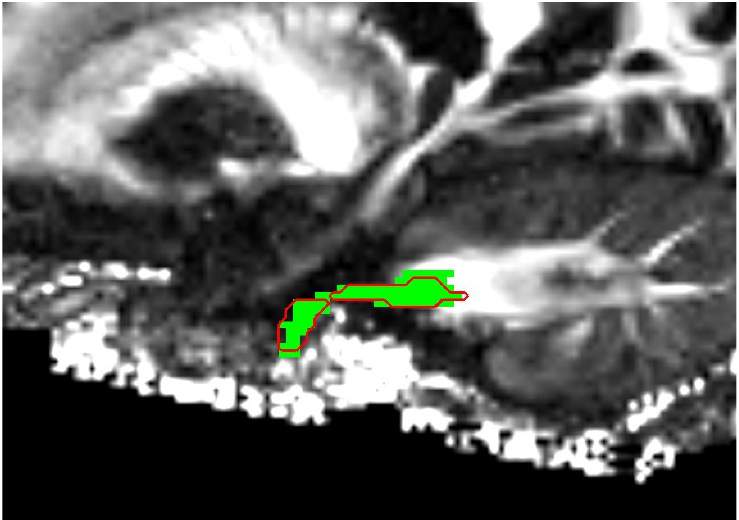}
			& \includegraphics[width=0.245\textwidth]{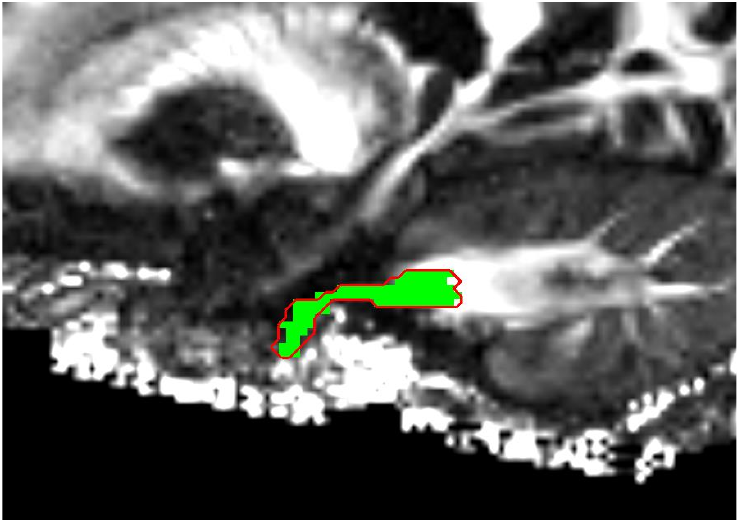} \\
			(C) w/o All &(D) CNTSeg-v2
			
	\end{tabular}}
	\caption{Comparison results on components of CNTSeg-v2: reconstruction of CN V on HCP subject \#116726. Green shows the reference tract and red shows the segmentation of the respective method.}\label{fig:ABLATION}
\end{figure}
\subsection{Ablation Study}
\subsubsection{Effectiveness on Components of CNTSeg-v2}
This section conducts an ablative study to investigate the contributions of different components in our CNTSeg-v2, consisting of the Arbitrary-Modal Collaboration Module (ACM) and Deep Distance-guided Multi-stage (DDM) decoder. For the fairness of the experiment, we removed these modules separately while maintaining the overall framework. All experiments were validated with 5-fold cross-validation on the HCP dataset with 102 subjects. Table.~\ref{tab:ablation} gives the segmentation performance of our CNTSeg-v2 without ACM, DDM, and all of them, respectively. For convenience, we have taken the five pairs of CNs as a whole and given mean values. From Table.~\ref{tab:ablation}, it is obvious that better results in segmenting CNs tract are achieved after using ACM, DDM in our CNTSeg-v2 respectively, as reflected in the four evaluation metrics we used. Specifically, the Dice for our CNTSeg-v2 can be up to 73.16\%, reduced by 0.36\% in the absence of ACM, and then by 0.14\% in the absence of all of ACM and DDM. Furthermore, the qualitative assessment of the ablation experiments is given in Fig.~\ref{fig:ABLATION}. We select the subject \#111413 on HCP dataset and give the segmentation results for CN V with complex structure on a sagittal slice of FA images. We can see that our CNTSeg-v2 have good anatomical correspondence in the brain pool segment with each component, reducing false positives even in the brainstem.
\begin{figure}[t]
	\centering
	\includegraphics[width=0.245 \textwidth]{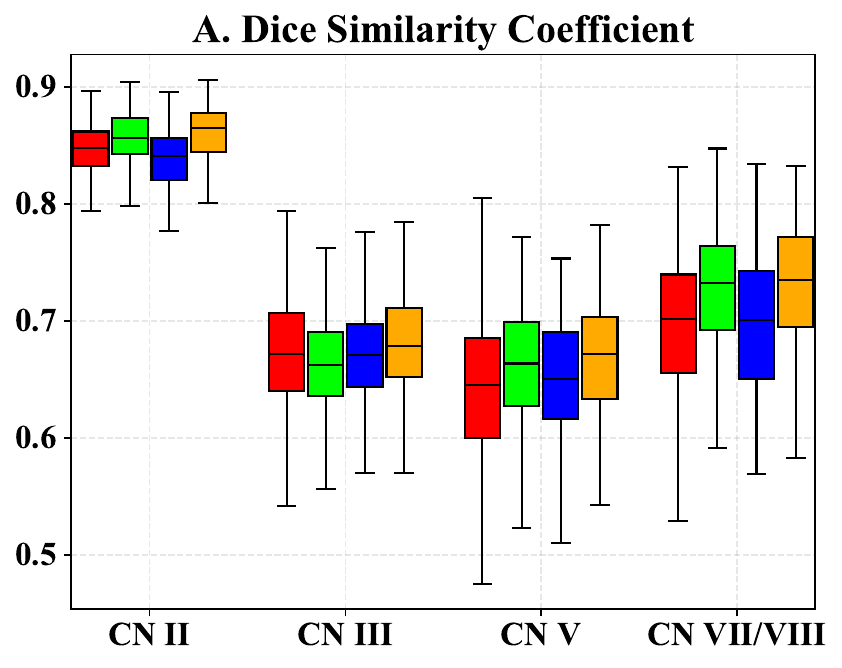}%
	\includegraphics[width=0.245 \textwidth]{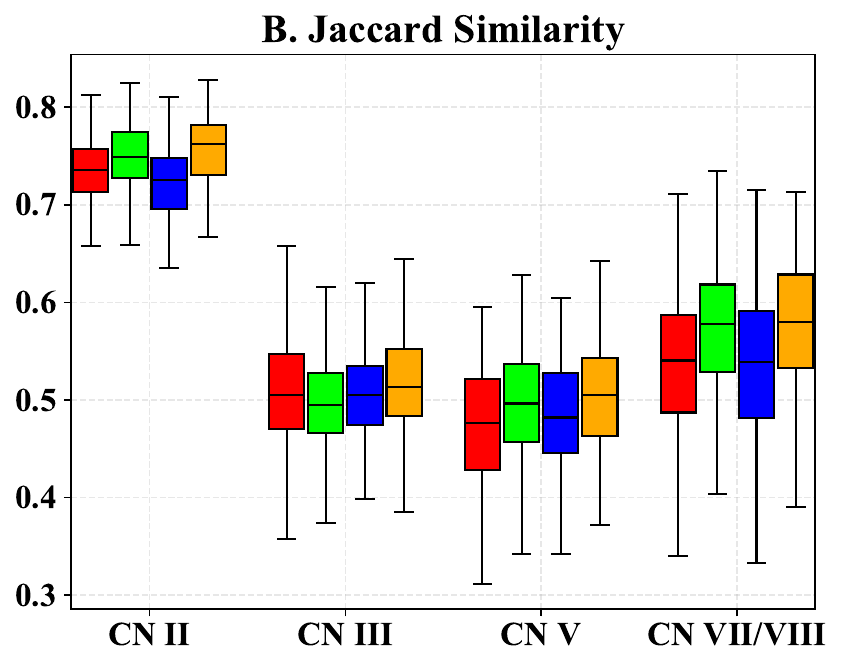}\\
	\includegraphics[width=0.245 \textwidth]{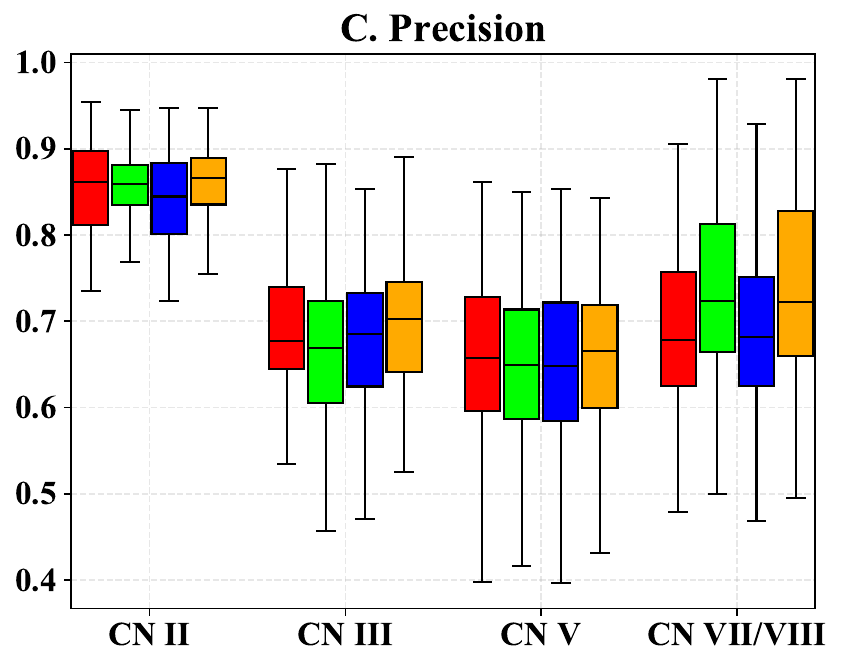}%
	\includegraphics[width=0.245 \textwidth]{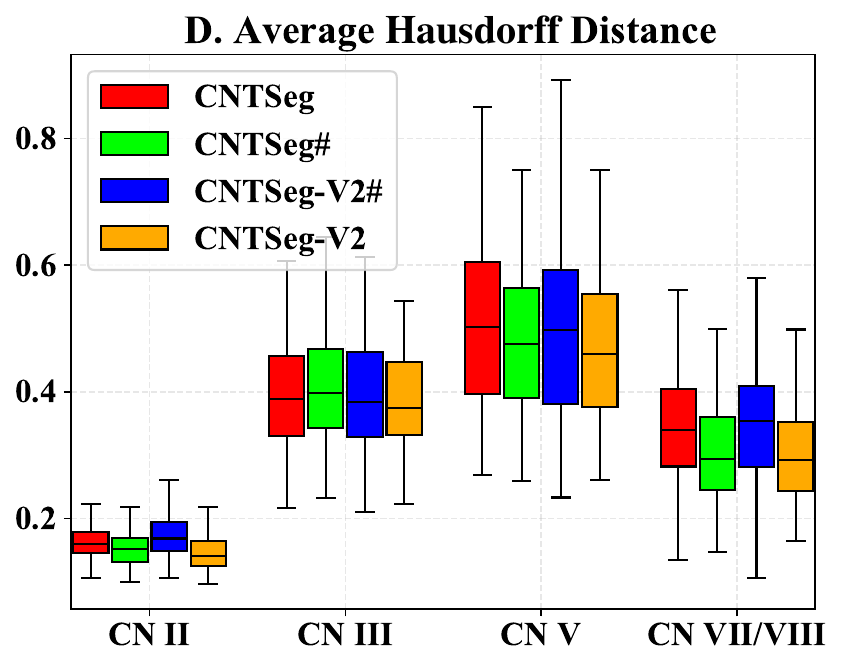}%
	\caption{Box-plots of segmentation results comparing CNTSeg-v2 and CNTSeg. CNTSeg\# represents models using \textit{T1w+T2w+FA+Peaks+DEC} combinations, and CNTSeg-v2\# represents model using \textit{T1w+FA+Peaks} combinations.}
	\label{fig:V1V2}
\end{figure}
\subsubsection{vs. CNTSeg}
CNTSeg was proposed in our previous work~\cite{xie2023cntseg}, which is an initial exploration of CNs tract segmentation using fully convolutional neural networks by using \textit{T1w+FA+Peaks} combinations. In this work, we make the first attempt to address the arbitrary cross-modal fusion model CNTSeg-v2 for CNs tract segmentation. To verify the segmentation performance of CNTSeg-v2 and CNTSeg, we compared them using different multimodal combinations: i) CNTSeg uses the original input modality, i.e., \textit{T1w+FA+Peaks} combination; ii) CNTSeg\# uses all modalities, i.e., \textit{T1w+T2w+FA+Peaks+DEC} combination; iii) CNTSeg-v2\# miss the T2w and DEC modality, i.e., \textit{T1w+FA+Peaks} combination; iv) CNTSeg-v2 uses the four modalities as auxiliary modalities, i.e., \textit{T1w+T2w+FA+Peaks+DEC} combination. Fig.~\ref{fig:V1V2}, we can see that CNTSeg-v2\# achieves higher performance in all four metrics compared to CNTSeg and CNTSeg-v2 achieves higher performance in all four metrics compared to CNTSeg\# under the same conditions. It is worth noting that the segmentation performance of CNTSeg decreases with an increasing number of modalities, suggesting that CNTSeg is sensitive to the settings of specific modalities, whereas CNTSeg-v2 demonstrates greater generalizability. As can be seen in Fig.~\ref{fig:SOTA} and Fig.~\ref{fig:mdm2}, our CNTSeg-v2 are significantly fewer false-positive fibers on the slice display compared to CNTSeg on HCP and MDM dataset. 
\begin{figure}[t]
	\centering
	\includegraphics[width=0.245 \textwidth]{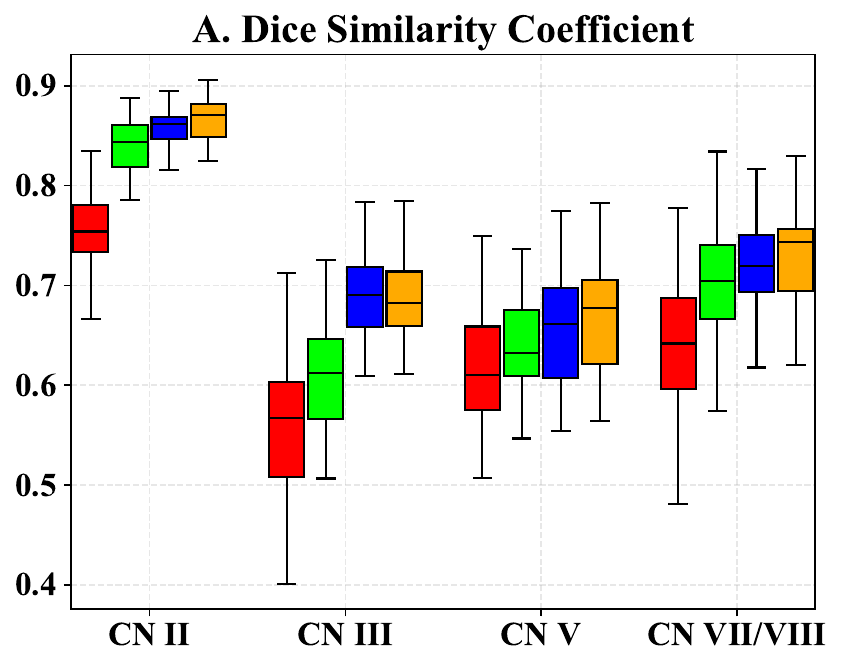}%
	\includegraphics[width=0.245 \textwidth]{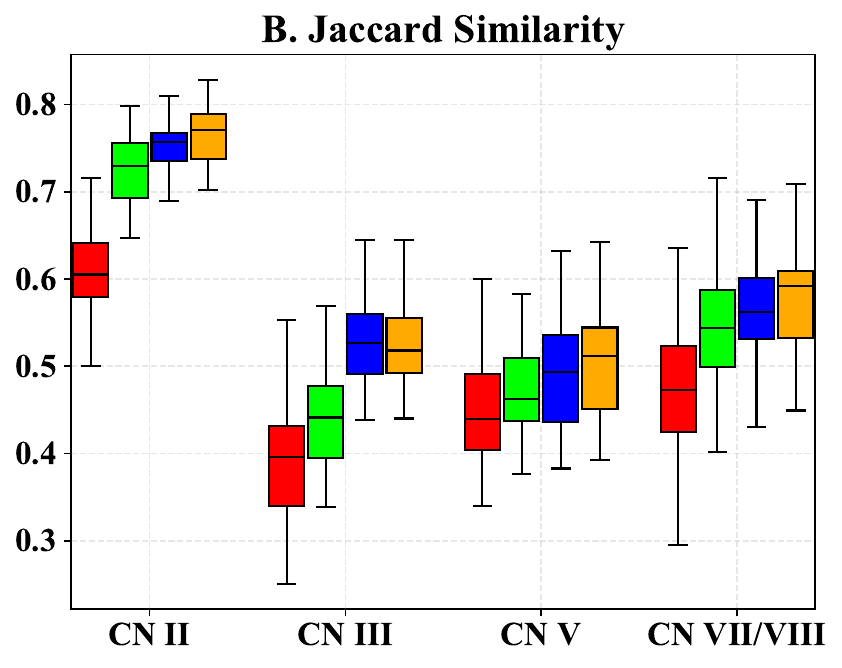}\\
	\includegraphics[width=0.245 \textwidth]{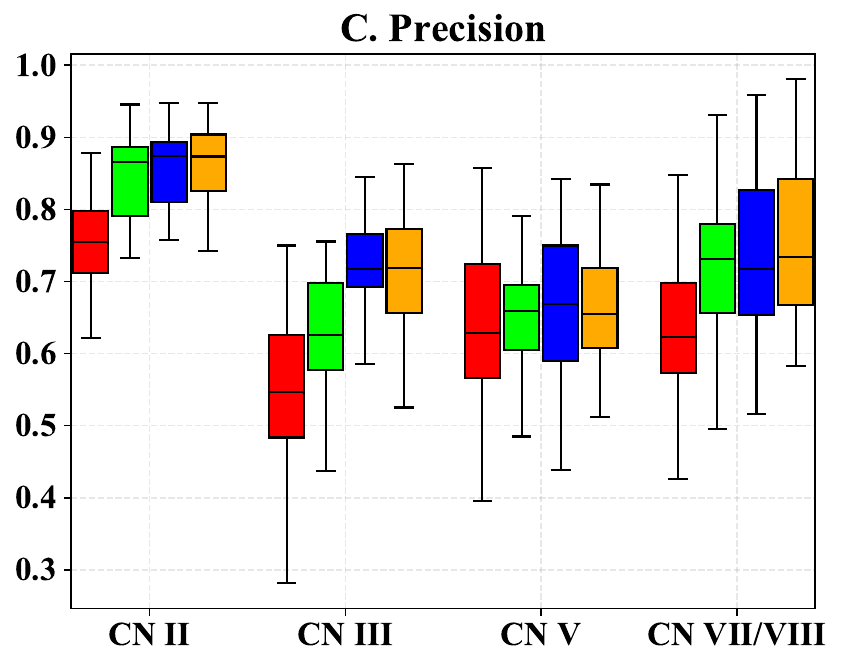}%
	\includegraphics[width=0.245 \textwidth]{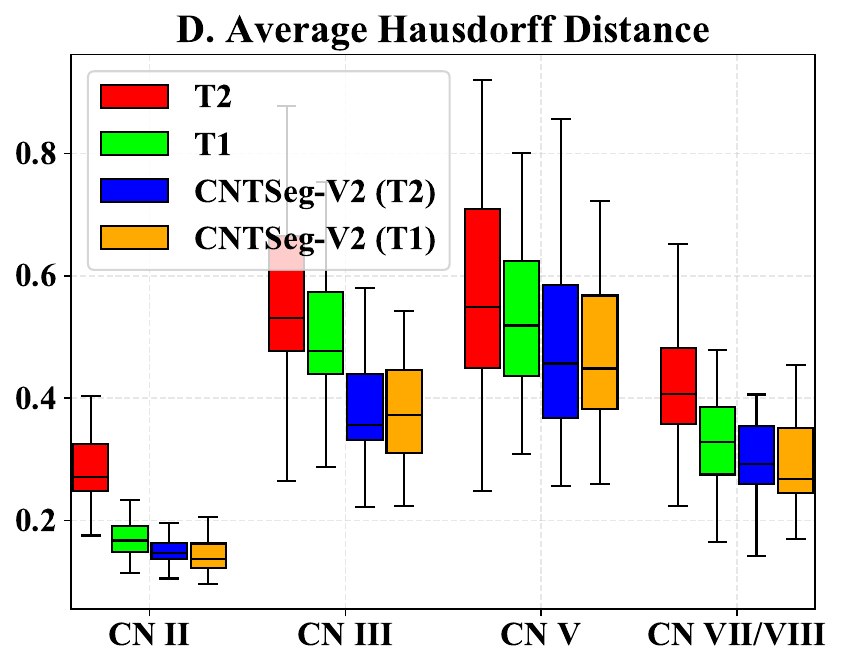}%
	\caption{Box-plots of segmentation results of CNTSeg-v2 using different primary modalities. T2w and T1w represent models using \textit{T2w} and \textit{T1w} single modality, respectively. CNTSeg-v2 (T2w) and CNTSeg-v2 (T1w) represent model using \textit{T2w} and \textit{T1w} as the primary modality, respectively.}
	\label{fig:T2}
\end{figure}
\subsubsection{Why T1w as the primary modality?}
Our CNTSeg-v2 model leverages the assumption that T1w images are essential for CNs tract segmentation, with other modalities used as supplementary information to enhance the model's segmentation capability. There are two primary reasons for this choice: simplicity in data acquisition and the significant contribution of T1w images to segmentation results.
Firstly, structural MRI, typically comprising both T1w and T2w images, is often the primary clinical examination for diagnosing cranial nerve disorders~\cite{diakite2025dual, zeng2025rgvpseg}. In contrast, diffusion MRI is more challenging to acquire due to stringent equipment parameter definitions, patient condition considerations, and inherent technical requirements.
Secondly, T1w images provide excellent contrast between different brain tissues. This ability to differentiate between brain nerves and surrounding tissues makes T1w images highly effective for CNs identification. As shown in Fig.~\ref{fig:T2}, the segmentation performance is given for model using only \textit{T2w} single modality, model using only \textit{T1w} single modality, CNTSeg-v2 using \textit{T2w} as the primary modality, and CNTSeg-v2 using \textit{T1w} as the primary modality, respectively. We can see that the CNTSeg-v2 using \textit{T1w} as the primary modality in five pairs of CNs for the highest segmentation performance.
Therefore, we select T1w images as the primary modality for CNs tract segmentation due to their straightforward acquisition process and substantial contribution to achieving superior segmentation results.

\section{Discussion}\label{sec:Discussion}

CNs tract segmentation is an important task in the clinical diagnosis of neurological diseases, and accurate identification of CNs can clearly demonstrate the spatial position of CNs in relation to the key surrounding tissues before surgery. Previously, we first explored deep learning network CNTSeg~\cite{xie2023cntseg} to achieve automatic and complete segmentation of five pairs of cranial nerves by introducing multimodal information. However, in clinical practice, acquiring a complete multimodal data is not always possible due to data corruption, varying scanning protocols, and unsuitable patient conditions. In this work, we propose a novel framework, called CNTSeg-v2, for arbitrary cross-modal CNs tract segmentation. Unlike CNTSeg, CNTSeg-v2 aims to enhance segmentation performance by integrating more modality-specific information. Experimental results demonstrate that CNTSeg-v2 outperforms existing neural segmentation methods. Even when employing the same modality combinations, CNTSeg-v2 consistently achieves optimal performance, highlighting its effective feature extraction from auxiliary modalities. Moreover, in scenarios where modalities are missing, CNTSeg-v2 demonstrates superior performance compared to RFNet and mmFormer, showcasing its capability to effectively complete cranial nerve segmentation tasks with only T1w images available.

However, there are some limitations to our study.
The first limitation is related to the reference data of our experiments. For generation of the reference data, we designed a semi-automatic generation method based on our previous CNs atlas. To reduce false positive fibers from the tracking method, we screened each pair of CNs using ROIs set by experts. Despite these efforts, the tracts we generated do not truly represent ground truth. Especially, in the brainstem, it is not possible to set the screening criteria manually, but only to roughly judge the region based on the course of streamlines, which inevitably leads to false positives. Despite these limitations, to the best of our knowledge, the generated reference data represents one of the best existing in-vivo approximations of known CNs anatomy. The second limitation is related to the specific datasets used in our experiments. We evaluate the proposed method on the HCP dataset and further validated it on the MDM dataset. However, the high image quality of these datasets complicates their direct application to clinical data and may necessitate fine-tuning of the model. Therefore, constructing a more generalized model that can be applied to low-quality clinical datasets is an urgent problem to address.

\section{Conclusion}\label{sec:Conclusion}
In this paper, we address arbitrary-modal CNs tract segmentation. We propose a universal model, CNTSeg-v2, for arbitrary-modal fusion, and introduce the Arbitrary-Modal Collaboration Module (ACM) to dynamically select complementary representations from other modalities, guided by the T1-weighted (T1w) modality for feature selection. Additionally, we design a Deep Distance-guided Multi-stage (DDM) decoder to supervise the segmentation results using a signed distance map. Extensive experimental results on the HCP dataset and the MDM dataset demonstrate that CNTSeg-v2 is more effective and achieves an improvement compared to SOTA methods in CNs tract segmentation.

\bibliographystyle{IEEEtran}
\bibliography{IEEEabrv,myreference}

\begin{thebibliography}{10}
\providecommand{\url}[1]{#1}
\csname url@samestyle\endcsname
\providecommand{\newblock}{\relax}
\providecommand{\bibinfo}[2]{#2}
\providecommand{\BIBentrySTDinterwordspacing}{\spaceskip=0pt\relax}
\providecommand{\BIBentryALTinterwordstretchfactor}{4}
\providecommand{\BIBentryALTinterwordspacing}{\spaceskip=\fontdimen2\font plus
\BIBentryALTinterwordstretchfactor\fontdimen3\font minus
  \fontdimen4\font\relax}
\providecommand{\BIBforeignlanguage}[2]{{%
\expandafter\ifx\csname l@#1\endcsname\relax
\typeout{** WARNING: IEEEtran.bst: No hyphenation pattern has been}%
\typeout{** loaded for the language `#1'. Using the pattern for}%
\typeout{** the default language instead.}%
\else
\language=\csname l@#1\endcsname
\fi
#2}}
\providecommand{\BIBdecl}{\relax}
\BIBdecl

\bibitem{yoshino2016visualization}
M.~Yoshino, K.~Abhinav, F.-C. Yeh, S.~Panesar, D.~Fernandes, S.~Pathak, P.~A.
  Gardner, and J.~C. Fernandez-Miranda, ``Visualization of cranial nerves using
  high-definition fiber tractography,'' \emph{Neurosurgery}, vol.~79, no.~1,
  pp. 146--165, 2016.

\bibitem{hodaie2010vivo}
M.~Hodaie, J.~Quan, and D.~Q. Chen, ``In vivo visualization of cranial nerve
  pathways in humans using diffusion-based tractography,'' \emph{Neurosurgery},
  vol.~66, no.~4, pp. 788--796, 2010.

\bibitem{jacquesson2019overcoming}
T.~Jacquesson, C.~Frindel, G.~Kocevar, M.~Berhouma, E.~Jouanneau, A.~Atty{\'e},
  and F.~Cotton, ``Overcoming challenges of cranial nerve tractography: a
  targeted review,'' \emph{Neurosurgery}, vol.~84, no.~2, pp. 313--325, 2019.

\bibitem{sultana2017mri}
S.~Sultana, J.~E. Blatt, B.~Gilles, T.~Rashid, and M.~A. Audette, ``Mri-based
  medial axis extraction and boundary segmentation of cranial nerves through
  discrete deformable 3d contour and surface models,'' \emph{IEEE Transactions
  on Medical Imaging}, vol.~36, no.~8, pp. 1711--1721, 2017.

\bibitem{xie2023cntseg}
L.~Xie, J.~Huang, J.~Yu, Q.~Zeng, Q.~Hu, Z.~Chen, G.~Xie, and Y.~Feng,
  ``Cntseg: A multimodal deep-learning-based network for cranial nerves tract
  segmentation,'' \emph{Medical Image Analysis}, vol.~86, p. 102766, 2023.

\bibitem{zolal2016comparison}
A.~Zolal, S.~B. Sobottka, D.~Podlesek, J.~Linn, B.~Rieger, T.~A. Juratli,
  G.~Schackert, and H.~H. Kitzler, ``Comparison of probabilistic and
  deterministic fiber tracking of cranial nerves,'' \emph{Journal of
  Neurosurgery}, vol. 127, no.~3, pp. 613--621, 2016.

\bibitem{jacquesson2019probabilistic}
T.~Jacquesson, F.~Cotton, A.~Atty{\'e}, S.~Zaouche, S.~Tringali, J.~Bosc,
  P.~Robinson, E.~Jouanneau, and C.~Frindel, ``Probabilistic tractography to
  predict the position of cranial nerves displaced by skull base tumors: value
  for surgical strategy through a case series of 62 patients,''
  \emph{Neurosurgery}, vol.~85, no.~1, pp. E125--E136, 2019.

\bibitem{xie2024anatomy}
L.~Xie, Q.~Zeng, H.~Zhou, G.~Xie, M.~Li, J.~Huang, J.~Cui, H.~Chen, and
  Y.~Feng, ``Anatomy-guided fiber trajectory distribution estimation for
  cranial nerves tractography,'' in \emph{2024 IEEE International Symposium on
  Biomedical Imaging (ISBI)}.\hskip 1em plus 0.5em minus 0.4em\relax IEEE,
  2024, pp. 1--5.

\bibitem{hu2024preoperative}
Q.~Hu, M.~Li, M.~Li, Q.~Zeng, J.~Yu, X.~Wang, Z.~Xia, L.~Xie, J.~Zhang,
  J.~Huang \emph{et~al.}, ``Preoperative diffusion tensor imaging:
  Fiber-trajectory-distribution-based tractography to identify facial nerve in
  vestibular schwannoma,'' \emph{Magnetic Resonance in Medicine}, vol.~92,
  no.~4, pp. 1755--1767, 2024.

\bibitem{he2021comparison}
J.~He, F.~Zhang, G.~Xie, S.~Yao, Y.~Feng, D.~C. Bastos, Y.~Rathi, N.~Makris,
  R.~Kikinis, A.~J. Golby \emph{et~al.}, ``Comparison of multiple tractography
  methods for reconstruction of the retinogeniculate visual pathway using
  diffusion mri,'' \emph{Human Brain Mapping}, vol.~42, no.~12, pp. 3887--3904,
  2021.

\bibitem{zeng2023automated}
Q.~Zeng, J.~Huang, J.~He, S.~Chen, L.~Xie, Z.~Chen, W.~Guo, S.~Yao, M.~Li,
  M.~Li, and Y.~Feng, ``Automated identification of the retinogeniculate visual
  pathway using a high-dimensional tractography atlas,'' \emph{IEEE
  Transactions on Cognitive and Developmental Systems}, vol.~16, no.~3, pp.
  818--827, 2024.

\bibitem{huang2022automatic}
J.~Huang, M.~Li, Q.~Zeng, L.~Xie, J.~He, G.~Chen, J.~Liang, M.~Li, and Y.~Feng,
  ``Automatic oculomotor nerve identification based on data-driven fiber
  clustering,'' \emph{Human Brain Mapping}, vol.~43, no.~7, pp. 2164--2180,
  2022.

\bibitem{2020Creation}
F.~Zhang, G.~Xie, L.~Leung, M.~Mooney, L.~Epprecht, I.~Norton, Y.~Rathi,
  R.~Kikinis, O.~Al-Mefty, N.~Makris, A.~Golby, and L.~O'Donnell, ``Creation of
  a novel trigeminal tractography atlas for automated trigeminal nerve
  identification,'' \emph{Neuroimage}, vol. 220, p. 117063, 06 2020.

\bibitem{zeng2021automated}
Q.~Zeng, M.~Li, S.~Yuan, J.~He, J.~Wang, Z.~Chen, C.~Zhao, G.~Chen, J.~Liang,
  M.~Li \emph{et~al.}, ``Automated facial--vestibulocochlear nerve complex
  identification based on data-driven tractography clustering,'' \emph{NMR in
  Biomedicine}, vol.~34, no.~12, p. e4607, 2021.

\bibitem{wasserthal2018tractseg}
J.~Wasserthal, P.~Neher, and K.~H. Maier-Hein, ``Tractseg-fast and accurate
  white matter tract segmentation,'' \emph{Neuroimage}, vol. 183, pp. 239--253,
  2018.

\bibitem{avital2019neural}
I.~Avital, I.~Nelkenbaum, G.~Tsarfaty, E.~Konen, N.~Kiryati, and A.~Mayer,
  ``Neural segmentation of seeding rois (srois) for pre-surgical brain
  tractography,'' \emph{IEEE Transactions on Medical Imaging}, vol.~39, no.~5,
  pp. 1655--1667, 2019.

\bibitem{xie2023deep}
L.~Xie, L.~Yang, Q.~Zeng, J.~He, J.~Huang, Y.~Feng, E.~Amelina, and M.~Amelin,
  ``Deep multimodal fusion network for the retinogeniculate visual pathway
  segmentation,'' in \emph{2023 42nd Chinese Control Conference (CCC)}.\hskip
  1em plus 0.5em minus 0.4em\relax IEEE, 2023, pp. 7946--7950.

\bibitem{dolz2017deep}
J.~Dolz, N.~Reyns, N.~Betrouni, D.~Kharroubi, M.~Quidet, L.~Massoptier, and
  M.~Vermandel, ``A deep learning classification scheme based on
  augmented-enhanced features to segment organs at risk on the optic region in
  brain cancer patients,'' \emph{arXiv preprint arXiv:1703.10480}, 2017.

\bibitem{mansoor2016deep}
A.~Mansoor, J.~J. Cerrolaza, R.~Idrees, E.~Biggs, M.~A. Alsharid, R.~A. Avery,
  and M.~G. Linguraru, ``Deep learning guided partitioned shape model for
  anterior visual pathway segmentation,'' \emph{IEEE Transactions on Medical
  Imaging}, vol.~35, no.~8, pp. 1856--1865, 2016.

\bibitem{li2021two}
S.~Li, Z.~Chen, W.~Guo, Q.~Zeng, and Y.~Feng, ``Two parallel stages deep
  learning network for anterior visual pathway segmentation,'' in
  \emph{Computational Diffusion MRI: International MICCAI Workshop, Lima, Peru,
  October 2020}.\hskip 1em plus 0.5em minus 0.4em\relax Springer, 2021, pp.
  279--290.

\bibitem{diakite2024lesen}
A.~Diakite, C.~Li, L.~Xie, Y.~Feng, H.~Han, P.~Dong, and S.~Wang, ``Lesen:
  Label-efficient deep learning for multi-parametric mri-based visual pathway
  segmentation,'' pp. 1--5, 2024.

\bibitem{zhang2023cmx}
J.~Zhang, H.~Liu, K.~Yang, X.~Hu, R.~Liu, and R.~Stiefelhagen, ``Cmx:
  Cross-modal fusion for rgb-x semantic segmentation with transformers,''
  \emph{IEEE Transactions on Intelligent Transportation Systems}, vol.~24,
  no.~12, pp. 14\,679--14\,694, 2023.

\bibitem{wang2023multistage}
J.~Wang, M.~Zhang, W.~Li, and R.~Tao, ``A multistage information complementary
  fusion network based on flexible-mixup for hsi-x image classification,''
  \emph{IEEE Transactions on Neural Networks and Learning Systems}, pp. 1--13,
  2023.

\bibitem{zhang2023delivering}
J.~Zhang, R.~Liu, H.~Shi, K.~Yang, S.~Rei{\ss}, K.~Peng, H.~Fu, K.~Wang, and
  R.~Stiefelhagen, ``Delivering arbitrary-modal semantic segmentation,'' in
  \emph{Proceedings of the IEEE/CVF Conference on Computer Vision and Pattern
  Recognition}, 2023, pp. 1136--1147.

\bibitem{ding2021rfnet}
Y.~Ding, X.~Yu, and Y.~Yang, ``Rfnet: Region-aware fusion network for
  incomplete multi-modal brain tumor segmentation,'' in \emph{Proceedings of
  the IEEE/CVF International Conference on Computer Vision}, 2021, pp.
  3975--3984.

\bibitem{zhang2022mmformer}
Y.~Zhang, N.~He, J.~Yang, Y.~Li, D.~Wei, Y.~Huang, Y.~Zhang, Z.~He, and
  Y.~Zheng, ``mmformer: Multimodal medical transformer for incomplete
  multimodal learning of brain tumor segmentation,'' in \emph{International
  Conference on Medical Image Computing and Computer-Assisted
  Intervention}.\hskip 1em plus 0.5em minus 0.4em\relax Springer, 2022, pp.
  107--117.

\bibitem{dolz2015fast}
J.~Dolz, H.-A. Leroy, N.~Reyns, L.~Massoptier, and M.~Vermandel, ``A fast and
  fully automated approach to segment optic nerves on mri and its application
  to radiosurgery,'' in \emph{2015 IEEE 12th International Symposium on
  Biomedical Imaging (ISBI)}, 2015, pp. 1102--1105.

\bibitem{wu2023hidanet}
Z.~Wu, G.~Allibert, F.~Meriaudeau, C.~Ma, and C.~Demonceaux, ``Hidanet: Rgb-d
  salient object detection via hierarchical depth awareness,'' \emph{IEEE
  Transactions on Image Processing}, vol.~32, pp. 2160--2173, 2023.

\bibitem{wang2020deep}
Y.~Wang, X.~Wei, F.~Liu, J.~Chen, Y.~Zhou, W.~Shen, E.~K. Fishman, and A.~L.
  Yuille, ``Deep distance transform for tubular structure segmentation in ct
  scans,'' in \emph{Proceedings of the IEEE/CVF Conference on Computer Vision
  and Pattern Recognition}, 2020, pp. 3833--3842.

\bibitem{yang2023exploring}
Y.~Yang, C.~Shan, F.~Zhao, W.~Liang, and J.~Han, ``On exploring shape and
  semantic enhancements for rgb-x semantic segmentation,'' \emph{IEEE
  Transactions on Intelligent Vehicles}, vol.~9, no.~1, pp. 2223--2235, 2024.

\bibitem{sotiropoulos2013advances}
S.~N. Sotiropoulos, S.~Jbabdi, J.~Xu, J.~L. Andersson, S.~Moeller, E.~J.
  Auerbach, M.~F. Glasser, M.~Hernandez, G.~Sapiro, M.~Jenkinson \emph{et~al.},
  ``Advances in diffusion mri acquisition and processing in the human
  connectome project,'' \emph{Neuroimage}, vol.~80, pp. 125--143, 2013.

\bibitem{van2013wu}
D.~C. Van~Essen, S.~M. Smith, D.~M. Barch, T.~E. Behrens, E.~Yacoub,
  K.~Ugurbil, W.-M.~H. Consortium \emph{et~al.}, ``The wu-minn human connectome
  project: an overview,'' \emph{Neuroimage}, vol.~80, pp. 62--79, 2013.

\bibitem{tong2020multicenter}
Q.~Tong, H.~He, T.~Gong, C.~Li, P.~Liang, T.~Qian, Y.~Sun, Q.~Ding, K.~Li, and
  J.~Zhong, ``Multicenter dataset of multi-shell diffusion mri in healthy
  traveling adults with identical settings,'' \emph{Scientific Data}, vol.~7,
  no.~1, p. 157, 2020.

\bibitem{tournier2019mrtrix3}
J.-D. Tournier, R.~Smith, D.~Raffelt, R.~Tabbara, T.~Dhollander, M.~Pietsch,
  D.~Christiaens, B.~Jeurissen, C.-H. Yeh, and A.~Connelly, ``Mrtrix3: A fast,
  flexible and open software framework for medical image processing and
  visualisation,'' \emph{Neuroimage}, vol. 202, p. 116137, 2019.

\bibitem{jeurissen2014multi}
B.~Jeurissen, J.-D. Tournier, T.~Dhollander, A.~Connelly, and J.~Sijbers,
  ``Multi-tissue constrained spherical deconvolution for improved analysis of
  multi-shell diffusion mri data,'' \emph{Neuroimage}, vol. 103, pp. 411--426,
  2014.

\bibitem{wang2021annotation}
S.~Wang, C.~Li, R.~Wang, Z.~Liu, M.~Wang, H.~Tan, Y.~Wu, X.~Liu, H.~Sun,
  R.~Yang \emph{et~al.}, ``Annotation-efficient deep learning for automatic
  medical image segmentation,'' \emph{Nature Communications}, vol.~12, no.~1,
  p. 5915, 2021.

\bibitem{xie2022semi}
L.~Xie, Z.~Chen, X.~Sheng, Q.~Zeng, J.~Huang, C.~Wen, L.~Wen, G.~Xie, and
  Y.~Feng, ``Semi-supervised region-connectivity-based cerebrovascular
  segmentation for time-of-flight magnetic resonance angiography image,''
  \emph{Computers in Biology and Medicine}, vol. 149, p. 105972, 2022.

\bibitem{diakite2025dual}
A.~Diakite, C.~Li, Y.~B.~M. Osman, Z.~Chen, Y.~Pan, J.~Zhang, T.~Tan, H.~Zheng,
  and S.~Wang, ``Dual-uncertainty guided multimodal mri-based visual pathway
  extraction,'' \emph{IEEE Transactions on Biomedical Engineering}, 2025.

\bibitem{zeng2025rgvpseg}
Q.~Zeng, L.~Yang, Y.~Li, L.~Xie, and Y.~Feng, ``Rgvpseg: multimodal information
  fusion network for retinogeniculate visual pathway segmentation,''
  \emph{Medical \& Biological Engineering \& Computing}, pp. 1--15, 2025.

\end{thebibliography}

\end{document}